\documentclass[conference]{IEEEtran}
\IEEEoverridecommandlockouts
\usepackage{cite}
\usepackage{amsmath,amssymb,amsfonts}
\usepackage{algpseudocode}

\usepackage{graphicx}
\usepackage{textcomp}
\usepackage{xcolor}
\usepackage{soul}
\usepackage{graphicx} % Required for including images
\usepackage{subcaption} % Required for creating subfigures
\def\BibTeX{{\rm B\kern-.05em{\sc i\kern-.025em b}\kern-.08em
    T\kern-.1667em\lower.7ex\hbox{E}\kern-.125emX}}

\usepackage{marginnote}
\usepackage[backgroundcolor=white,bordercolor=purple,linecolor=purple]{todonotes}

\newcommand{\MAJOR}[2]{{#2}}

\newcommand{\MAJOREND}{\color{black}}

\usepackage{xspace}

\usepackage{multirow}

\usepackage{array}
\usepackage{enumitem}
\usepackage{subcaption}

\newcommand{\NMAP}{Nmap\xspace}
\newcommand{\SES}{SES\xspace}

\newcommand{\APPR}{FISTS\xspace}

\definecolor{byzantine}{rgb}{0.74, 0.2, 0.64}

\begin{document}

\title{Field-based Security Testing of SDN configuration Updates}

% \author{\IEEEauthorblockN{Jahanzaib Malik, Fabrizio Pastore}
% \IEEEauthorblockA{\textit{University of Luxembourg} \\
% \textit{Snt Centre}\\
% Anonym \\
% Anonym}
% }

\author{Jahanzaib Malik, Fabrizio Pastore,~\IEEEmembership{Member,IEEE}
        % <-this % stops a space
\thanks{This paper was produced by Jahanzaib Malik, Fabrizio Pastore. They are affiliated with University of Luxembourg, SnT centre. E-mails: jahanzaib.malik@uni.lu fabrizio.pastore@uni.lu }% <-this % stops a space
\thanks{Manuscript received January 30, 2024; revised XXXX XX, XXXX.}}

\maketitle

\begin{abstract}
Software-defined systems revolutionized the management of hardware devices but introduced quality assurance challenges that remain to be tackled. \MAJOR{1.1}{For example, software defined networks (SDNs) became a key technology for the prompt reconfigurations of network services in many sectors including 
telecommunications, data centers, financial services, cloud providers, and manufacturing industry.}  
Unfortunately, reconfigurations may lead to mistakes that compromise the dependability of the provided services. 
\MAJOR{1.1}{In this paper, we focus on the reconfigurations of network services in the satellite communication sector, and target security requirements, which are often hard to verify;} for example, although connectivity may function properly, confidentiality may be broken by packets forwarded to a wrong destination. 
%Approaches for SDN security testing and the verification of SDN configurations are of limited usefulness in industrial contexts because they either target SDN component testing, not configuration testing, or make assumptions about the opensource nature of the network stack.
We propose an approach for FIeld-based Security Testing of SDN Configurations Updates (\APPR). First, it probes the network before and after configuration updates.
%pplying unsupervised machine learning algorithms to the network packets observed in the field after probing the network before and after configuration updates.
Then, using the collected data, it relies on unsupervised machine learning algorithms to prioritize the inspection of suspicious node responses, after identifying the network nodes that likely match across the two configurations.
Our empirical evaluation has been conducted with network data from simulated and real SDN configuration updates for our industry partner, \MAJOR{3.1}{a world-leading satellite operator.} Our results show that, when combined with K-Nearest Neighbour, \APPR leads to best results (up to 0.95 precision and 1.00 recall). Further, we demonstrated its scalability.
\end{abstract}




% !TEX root =  MAIN.tex

\section{Introduction}

Communication sectors, \MAJOR{5.6}{including satellite communication (SATCOM~\cite{SATCOM}) and terrestrial telecommunication (TELCO),} are critical for our daily life and thus need to continuously evolve to enable new services for demanding customers while containing costs. 
Example SATCOM services include fast internet for aviation~\cite{aviation}, navy~\cite{maritime}, and land~\cite{fastInternet}, disaster recovery~\cite{disaster}, connection of remote communities~\cite{remote}, DTH broadcasting~\cite{broadcasting}, IoT connectivity~\cite{IoT}.

What enables the delivery of high-quality (e.g., high speed, low delay) services over an infrastructure that is costly to grow (e.g., because of costs related to satellite deployment) is the quick reconfiguration of the infrastructure, which enables reusing components (e.g., satellites) for different purposes. In the communication sector, a key technology to achieve such reconfiguration capabilities are Software Defined Networks (SDNs).
This is the case for our industry partner \SES, one of the world leading satellite operators. 

SDNs enable dynamic and programmatic network configuration. They disassociate the delivery of network packets (data plane) from the routing process (control plane). Specifically, SATCOM companies rely on software-defined wide area networks (SD-WAN), which means that SDN concepts are adopted to reconfigure a Wide Area Network (WAN).

In SDN-WANs, re-configurations are \MAJOR{5.5}{frequent; they may concern both the control and the application plane~\cite{SWAN,electronics13153011}}. Unfortunately, the presence of multiple applications (e.g., firewall, router) interacting with the data packets flowing through the SDN may lead to inconsistent or conflicting configurations, \MAJOR{1.3}{as reported in empirical studies\cite{al2021migrating,yungaicela2024misconfiguration}.} 
We refer to such situations with the generic term of \emph{SDN misconfigurations}.
Although misconfigurations may break several types of system requirements (e.g., functional, security, robustness), in this paper we focus on non-functional requirements, specifically, security requirements, because they are subtle to detect. Indeed, while violations of functional requirements might be easily discovered (e.g., the service is not delivered), violations of security requirements may not have any visible effect. For example, a packet might be routed through an unsecure component before correctly reaching the final destination thus violating confidentiality requirements without triggering any alarm. 

Other sectors embraced the transition to software-defined systems only recently (e.g., software-defined vehicles~\cite{SDVehicles}), and thus may be affected by the challenges introduced by simplified reconfigurations in the coming future.

% %The growing popularity of SDNs has lead to a number of approaches testing SDN security. Unfortunately, several approaches test the
% Testing SDN configurations is not a new problem.

% Unfortunately,  Finally, we assume that all the SDN components can be trusted because provided by major vendors; in other words, we are not interested in testing the effect of a malicious component in a SDN, which is the focus of most of the literature on SDN security testing. One possible extension is to test the effects of reconfiguration on customer’s services (e.g., check if a server running on the customer’s network can be attacked just because the customer has asked to open a certain range of ports).
% Examples of the misconfigurations mentioned above may include (1) a URL that is whitelisted to ensure software updates but is blocked by IP filtering configurations or (2) flow rules bringing data to unsecure or inappropriate networks (e.g., the data of a customer is partially redirected to the network of another customer). Therefore, it is important to ensure that an updated configuration leads to the desired effect. 
% Background and Related work

Ensuring that SDN configurations match service requirements is a well-known engineering problem. 
According to a recent survey, most of the approaches targeting such objective rely on model-based verification~\cite{Rojas2018}. They either rely on model-checking strategies to identify conflicts (e.g., NICE~\cite{NICE}) or derive inputs from an SDN configuration model to test if the SDN leads to the expected outcomes (e.g., TASTE~\cite{TASTE}, ATPG~\cite{ATPG}, and BUZZ~\cite{BUZZ}). 
%(e.g., URL can be accessed) 
The main limitation of such approaches is that they rely on a model of the intended SDN configuration, which should be either manually produced by an engineer or derived from the configurations provided to the SDN management software. Both these two cases are expensive, indeed the first implies that engineers manually produce a model, which is time consuming, the second implies the development of a tool that is tightly integrated with the SDN management software (e.g., OpenDaylight~\cite{opendaylight}), which is hard to achieve because, in industry, SDN management software is often proprietary (e.g., Versa Secure SD-WAN~\cite{VERSA}) and not developed by the company delivering services. Further, companies often change software supplier based on business decisions thus making in-house development of such tools financially unviable as it requires R\&D expertise, often off-loaded to other stakeholders. 

%software (e.g., it should process Versa file format); such integration may be expensive to maintain. In addition to that, a problem that may affect model-based verification approaches relying on model checking is that they require detailed modelling(insight) of the system (what is not modelled is not tested) and furthermore, may not detect problems visible only with the actual implementation (e.g., because of delays during transmission). 

Testing approaches not relying on SDN models might be more applicable, in practice. Such approaches are often referred in the literature as fuzz testing approaches or fuzzers. Known SDN fuzzers are DELTA\cite{Lee2020} and BEADS~\cite{Jero}, but they aim at identifying faults in SDN applications or SDN control plane components, not detecting misconfigurations. Testing SDN components, is typically out of the interests of service providers who rely on third party suppliers that provide contractual guarantees about the security of the SDN components in use. 
Alternatively, field-based testing approaches enable testing systems deployed in the production environment~\cite{Bertolino2021} and, therefore, enable configuration testing. However, the number of field-based testing techniques targeting software security is limited, seven out of 80 papers appearing in a recent survey~\cite{Bertolino2021}. Further, they test the security of a single service, not the whole SDN system.
%Indeed, four approaches address problems in online service compositions~\cite{bertolino2007plastic,bertolino2011enhancing,de2011role,zhang2004approach}, one approach targets only integer overflows~\cite{hui2016runtime}, two approaches~\cite{dai2010configuration,dai2012confu}, in addition to test configurations of a single software component, concern offline testing (i.e., they test sibling processes with modified configurations), which is infeasible in our context.
To summarize, the automated identification of SDN misconfigurations leading to security issues remains an open problem.

We propose FIeld-based Security Testing of SDN Configurations Updates (\APPR), a field-based testing approach that works in four steps. In the first step, we scan the SDN network, before and after a configuration change, with a predefined set of data packets (e.g., the ones generated by the NMAP security scanner~\cite{NMAP}) and collect response data to determine the state (e.g., open ports) of the reachable hosts\footnote{We use the term host to refer to a generic machine on the network, whether it is a server, router, switch, or end-user terminal.}. In the second step, we automatically match the hosts identified with the two network scans (e.g., a same host changed IP) to determine changes in their state. \MAJOR{3.2}{In the third step, we prioritize the inspection of the scanned hosts by leveraging the results of anomaly detection algorithms. Specifically, our approach can leverage well known anomaly detection algorithms such as isolation forest~\cite{IF} and local outlier factor~\cite{LOF}; further, we propose two solutions for the prioritization of anomalies  that are based on KNN~\cite{RamaswamyKNN}, HAC~\cite{King:2014} and k-means~\cite{mcqueen1967smc}.
As a result, the prioritized list includes items likely affected by vulnerabilities on top; note that in this context, a vulnerability is a specific host state (e.g., port 8080 is reachable) due to an SDN misconfiguration (e.g., a stateful firewall SDN application had not been set up).} In the fourth step,  engineers inspect the prioritized list of hosts and stops when the list does not present any more vulnerabilities; we empirically determined a threshold for the number of consecutive false positives to be observed before stopping. \MAJOR{4.1}{Alternatively, engineers can inspect all the hosts in the prioritized order, to avoid missing any vulnerability, but prioritizing the inspection of vulnerable ones.}

We conducted an empirical assessment with different datasets of network updates to determine the best configuration for \APPR by comparing results achieved with and without pruning, along with different anomaly detection algorithms. Further, we demonstrate the accuracy of our host matching component. \MAJOR{3.3}{Last, we report on the scalability of our network scanning method, and the selected anomaly detection algorithms, by reporting on the time required to monitor and process 400 network nodes.} 

\MAJOR{4.1,5.6}{To summarize, our contributions are:
\begin{itemize}
\item \APPR, the first approach for field-based testing of security properties that combines network scanning and anomaly detection; the approach presents two modalities, one to support the inspection of a subset of hosts only, one to inspect all the hosts.
\item SKNN, an approach for the prioritization of anomalies based on the KNN algorithm;
\item An approach for the prioritization of anomalies based on clustering algorithms, which has been applied to HAC (hereafter, SHAC) and K-means (hereafter, SKM);
\item An extensive empirical evaluation conducted on 220 datasets with synthetic and real data.
\end{itemize}}

%the accuracy The best approach is based on KNN, s provides the best results in terms of precision, recall, and for different  stopping criteria. Further, we assessed the proposed approach with historical data concerning our industry partner and available online. Finally we report on the scalability of the approach. Our empirical results show that the best performing algorithm is isolation forest, the approach can identify 60\% of the vulnerabilities in realistic scenarios with a minimal effort (25 hosts inspected), and the execution of the approach takes only XX seconds per host.

This paper proceeds as follows: 
Section~\ref{sec:background} introduces background and related work.
Section~\ref{sec:approach} describes the proposed approach.  
Section~\ref{sec:empirical} reports on the results of our empirical assessment.  
Section~\ref{sec:conclusion} concludes the paper.

% !TEX root =  MAIN.tex

\section{Background and Related Work}
\label{sec:background}

\MAJOR{1.10, 2.1}{By identifying anomalies in network data, we aim at detecting configuration updates that are faulty and, consequently, introduce vulnerabilities. In this section, we present related (verification of SDN configuration) and background approaches (network scanning with NMAP, data clustering, and anomaly detection).} 

\subsection{Verification of SDN configurations}
\label{sec:related}

Approaches for the verification of SDN configurations rely on either model analysis or testing~\cite{Rojas2018}, as explained below.

Model-based approaches can be used to verify that SDN systems implement the provided configuration either by relying on model-checking strategies to identify conflicts \cite{Rojas2018}. 
\MAJOR{1.3}{For example, Saied et al. automatically detect and correct misconfigurations in SDN switches~\cite{saied2020formal}. Initially, they collect configuration data to construct formal models representing both intended and actual network behaviors using network switch flows. Then, they rely on model checking to identify discrepancies between these models, assess their impact, and align actual operations with intended configurations. 
%Corrections are validated through re-evaluation, ensuring the network's integrity, with continuous monitoring to detect and rectify new discrepancies as they arise. 
Sadauoi et al., detect misconfiguration  in OpenFlow flow rules utilizing a flow table decision diagram (FtDD) created from the flow rules of all the switches on the network~\cite{saadaoui2019automated}. They parse all possible paths that a packet could take in the given network topology to detect conflicting rules. Another approach relies on flow rules to create a binary decision diagram (BDD) that is processed by relying on Computational Tree Logic (CTL) queries to detect intra-rule misconfigurations~\cite{FlowTable_misconfigurations_SDN}. Pan et al., instead, present a dedicated algorithm that relies on intervals derived from flow rules to check for redundant rules and minimize the total number of rules~\cite{pan2022misconfiguration}.} Le Brun et al., instead, automatically derive inputs (e.g., network packets) from the configuration itself (e.g., to request a whitelisted URL) to test if the SDN leads to the expected outcomes (e.g., URL can be accessed)~\cite{Lebrun2014}.

The main limitation of the approaches above is that they rely on a model of the system, which should be either manually produced by the engineer or derived from the configurations provided to the SDN software. Both cases are expensive, indeed, the former directly requires engineers to produce a model, the latter requires the tool to be tightly integrated with the SDN software (e.g., it should process the proprietary format of commercial tools such as Versa~\cite{VERSA}); such integration might be expensive to maintain. In addition to that, a problem that may affect model-based verification approaches relying on model checking is that they require detailed modelling of the system (what is not modelled is not tested) and furthermore, may not detect problems visible only when the system is executed (e.g., because of delays during transmission). %According to a recent survey, verification approaches are the majority.

Because of the above, automated testing approaches not relying on SDN modelling might be more adequate to address our problem. In the context of SDN testing, researchers have explored the use of fuzz testing~\cite{manes2019art}, \MAJOR{1.10}{which led to the development of DELTA~\cite{Lee2020}, BEADS~\cite{Jero2017}, FragScapy~\cite{Fragscapy}, and AFLNET~\cite{Pham2020}. Although such fuzzers} do not directly address the problem of testing SDN misconfigurations, they might be leveraged to generate traffic in place of NMAP. DELTA generates anomalous traffic by altering valid data packets. Most of the vulnerabilities detected by DELTA result from the modification of control flow messages, which is out of scope for our project; however, DELTA includes a Host agent that is useful for launching some attacks initiated by hosts (it generates network flows by creating new TCP connections or by using existing utilities, such as Tcpreplay). BEADS aims at discovering failures that may be triggered by the presence of malicious SDN components (applications or controllers) and therefore its applicability is out of scope since, as mentioned in the Introduction section, SATCOM providers assume that all the components installed on a SATCOM SDNs can be trusted and they are not interested in testing them. For the same reason, approaches for the identification of adversarial data planes are out of scope as well~\cite{Black2021}.
FragScapy is a command-line tool that can be used to generate 'fragroute-like' tests using Scapy~\cite{Scapy}. FragScapy is useful to collect network data that can't be collected by NMAP in the presence of firewalls, but, alone, does not enable detecting misconfigurations. AFLNet targets implementations of network protocols thus targeting a problem different than ours.

In addition to FragScapy, other tools can generate test traffic and might be considered in the future to complement NMAP in \APPR; we describe them in the following. ATPG~\cite{Zeng2012} generates packets to test network availability and verify packet latency along with the consistency of the data plane with respect to configuration specifications; BUZZ~\cite{Fayaz2016} focuses on policy implementations; Monocle \cite{perevsini2015monocle} checks inconsistencies in the data plane; sPing~\cite{tseng2017sping} relies on packet injection to discover network loops, black holes, and link layer information; RuleScope~\cite{bu2016every} inspects SDN forwarding; SDN traceroute~\cite{agarwal2014sdn} is a packet-tracing tool for measuring paths in SDN networks; sTrace~\cite{wang2016tool-strace} is a packet-tracing tool for large, multi-domain SDN networks; FLOWGUARD~\cite{hu2014flowguard} detects firewall policy violations; Libra~\cite{zeng2014libra} detects loops, black holes, and other reachability failures; HSA (Header Space Analysis)~\cite{HSA2012} is a protocol-agnostic framework to identify reachability failures, forwarding loops, and traffic isolation; FlowTest and GraphPlan \cite{fayaz2014testing} tackle stateful and dynamic data plane functions (DPFs), and policy requirements. \MAJOR{5.7}{To conclude, testing approaches aim to detect problems that are complementary to our (e.g., high latency, network loops); however, some of them might be leveraged to collect additional data to extend \APPR in the future. Feasible options include FragScapy to complement NMAP limitations, and collect latency data for anomaly detection; however, NMAP remains the best choice for monitoring available ports.}

\MAJOREND{}

% !TEX root =  MAIN.tex
%\section{Background}
%\label{sec:background}

\subsection{Network Scanning with NMAP}
\label{back:nmap}
Network Mapper (NMAP) is an open-source network scanning tool, which is the industry-wide standard for exploring networks. NMAP offers various scanning techniques and strategies which enable professionals to discover network hosts and identify running services. 
%It supports a wide range of scanning options, a subset of which is presented in Table~\ref{table:NMAP:scans}.
%including TCP, UDP, SYN, and ICMP scans, allowing users to customize their scans based according to their needs.
For every scanned host, NMAP reports the state of the probed ports; the possible cases are shown in Table~\ref{table:NMAP:portstates}.

\MAJOR{1.4}{Since service providers (e.g., SATCOM companies) may rely on proprietary SDN protocols that may change over time, we test the SDN in a black-box manner. Specifically, we do not rely on the NMAP options for OpenFlow (e.g., openflow-info script).}

%Further, through its Scripting Engine (NSE), \NMAP can be extended to perform vulnerability detection and exploitation.

%\input{tables/NMAP_SCAN.tex}

% !TEX root =  ../MAIN.tex

% \begin{table}[tb]
% \caption{Port states determined by \NMAP .}
% \footnotesize
% \begin{tabular}{|@{}
% p{18mm}
% @{\hspace{1mm}}|p{13mm}
% @{\hspace{1mm}}|p{5.3cm}@{\hspace{1mm}}|} \hline 

% \textbf{Port state}&\textbf{Scan Type}&\textbf{Description}\\ \hline 

% open& All& An application is actively accepting TCP connections, UDP datagrams, or SCTP associations on this port.  \\ \hline 
% closed& All& The port is accessible (it receives and responds to \NMAP), but there is no application listening on it. \\ \hline 
% filtered& All& \NMAP cannot determine whether the port is open because packet filtering (e.g., firewall) prevents responses to reach \NMAP.\\ \hline 
% unfiltered& ACK& The port is accessible, but \NMAP is unable to determine whether it is open or closed.\\ \hline  
% open$|$filtered&UDP, IP, FIN, NULL, Xmas& \NMAP is unable to determine whether a port is open or filtered because the selected scan type does not enable getting any response from the port.\\ \hline 
% closed$|$filtered& IP ID idle&\NMAP is unable to determine whether the port is closed or filtered.\\ \hline

% \end{tabular}

% \label{table:nmap:ports}
% \end{table}%

\begin{table}[tb]
\caption{Port states determined by \NMAP .}
\label{table:NMAP:portstates}

\scriptsize
\centering
\begin{tabular}{p{0.59in}p{0.49in}p{1.8in}}

\hline 
\textbf{Port State} & \textbf{Scan Type}       & \textbf{Description}                                                                                                                                              \\ \hline 
Open                & All                      & An application is actively accepting TCP connections, UDP datagrams, or SCTP associations on this port.                                                           \\ \hline
Closed              & All                      & The port is accessible (it receives and responds to \NMAP), but there is no application listening on it.                                           \\ \hline
Filtered            & All                      & NMAP cannot determine whether the port is open because packet filtering (e.g., firewall) prevents responses to reach NMAP.                                        \\ \hline
Unfiltered          & ACK                      & The port is accessible, but \NMAP is unable to determine whether it is open or closed                                                              \\ \hline
Open$|$Filtered       & UDP, IP, FIN, NULL, Xmas & NMAP is unable to determine whether a port is open or filtered because the selected scan type does not enable getting any response from the port. \\ \hline
Closed$|$Filtered     & IP ID idle               & NMAP is unable to determine whether the port is closed or filtered.   
\\ \hline 
\end{tabular}
\end{table}

% Nmap also enables to identify the operating system(along with their versions) and open ports for a given IP. 
% In terms of its utility it can easily be integrated with any system using its Nmap Scripting Engine (NSE), which empowers developers and security professionals to use its capabilities in creative ways to ensure security of organizations. Moreover, it can also be used as a command line utility (CLI) and with graphic user interface (GUI) know as Zenmap. Give the fact its open-source, developers are regularly trying to improve it and equip it with more sophisticated scans and techniques which in turn produce better results and insights for security professionals.

\subsection{Data clustering}
\label{back:clustering}

Data clustering involves the partitioning of a dataset into groups, or clusters, based on the similarity among the data points within each group~\cite{Berkhin2006,TNN:SURVEY:2005,DataClusteringBook:2013}. 
%The objective of clustering is to identify patterns or relationships in the data without prior knowledge of the groupings. 
Clustering algorithms differ for the strategy adopted to partition the dataset; their performance depends on the characteristics of the dataset~\cite{TNN:SURVEY:2005}.
Among traditional clustering algorithms~\cite{Survey:2015}, hierarchical algorithms (e.g., HAC~\cite{King:2014} and BIRCH~\cite{BIRCH}) perform well when clusters have a tree-shaped form, partition-based algorithms (e.g., K-means~\cite{mcqueen1967smc} and PAM~\cite{PAM}) assume that clusters have the form of hyper-spheres, density-based algorithms (e.g., DBSCAN~\cite{DBSCAN}, OPTICS~\cite{OPTICS}, and Mean Shift~\cite{MeanShift}) do not make assumptions on the clustering shape but assume that points belonging to a same dense region belong to the same cluster. Below, we detail the approaches considered in this paper (i.e., HAC and K-means).
%K-nearest neighbor (KNN), Local Outlier Factor (LOF), Isolation Forest (IF) and K-means).

% HAC
Hierarchical agglomerative clustering (HAC) is a bottom up approach in which each data point starts in its own cluster; it iteratively merges pairs of clusters thus producing a dendrogram. The input of HAC is a matrix capturing the distance between every pair of data point. Grouping aims at minimizing one objective function; in our work we rely on the \emph{ward} linkage method, which minimizes the variance of the clusters being merged \cite{mullner2011modern}.

% The grouping that occurs at each step aims to minimize an objective function.
% %the impairment of the optimum value of an objective function. 
% In HAC, widely adopted objective functions, which we use in our work, are 
% \ASEnnn{the error sum of squares within clusters (i.e.,Ward’s linkage method~\cite{Ward}), to help minimize within-cluster variance, and the average of distances between all pairs of elements belonging to distinct clusters (i.e.,average linkage~\cite{UPGMA}), to help maximize diversity among clusters.}

% k-means
K-means is the most popular clustering algorithm; 
%unsupervised machine learning technique used for partitioning a dataset into distinct groups or clusters based on similarity patterns among data points. 
it works by iteratively assigning data points to the nearest cluster centroid and recalculating the centroids until convergence. The objective is to minimize the within-cluster sum of squares, making data points within the same cluster as similar as possible while maximizing dissimilarity between clusters. 
%K-means is widely applied in various fields, including data analysis, image segmentation, and customer segmentation.
%offering a simple yet effective method for discovering underlying structures and patterns within large datasets, thereby aiding in data exploration and decision-making processes.

% comparing Agglomerative (HAC) and K-means
Both HAC and K-means do not automatically determine how many clusters should be generated. To determine the number of clusters that best separate our dataset we rely on Silhouette analysis~\cite{ROUSSEEUW198753}, which is a state-of-the-practice solution for this purpose.
Precisely, we execute our clustering algorithms multiple times, with a number of clusters ranging from 2 to N (in our experiments we set N to 50).
According to Silhouette analysis we select the clustering output having the largest number of clusters with a max Silhouette coefficient above the Silhouette score. The Silhouette coefficient is computed for every data point in a cluster as the difference between the average distance from the data points within a cluster and the data points in the closest cluster normalized by the largest of the two. The Silhouette score is the average Silhouette index across all the datapoints in a cluster.

\subsection{Anomaly detection}
\label{back:anomalyDetection}
Anomaly detection refers to the process of identifying and flagging data points that significantly differ from the expected or normal behavior within a given dataset~\cite{survey:anomaly:2009,OutlierSurvey}.
\MAJOR{3.2}{Specifically, we aim to identify outliers within tabular data. A recent survey classifies fundamental outlier detection approaches into
nearest-neighbor-based, projection-based, and clustering-based~\cite{OutlierSurvey}. 
In this paper, we consider one representative for each category, which are Local Outlier Factor (LOF)~\cite{LOF} and 
Isolation Forest (IF)~\cite{isolation_forest}, for the first two categories. We selected these two algorithms because they are accurate (as reported in the survey) and largely adopted by data analysts (e.g., included in well-known libraries). 
Clustering-based methods, instead, rely on popular clustering algorithms (e.g., k-means, see Section~\ref{back:clustering}) but differ for their rationales, which are: (a) anomalies do not belong to any cluster, (b) anomalies are away from the cluster centroids, (c) anomalies belong to small or sparse clusters~\cite{survey:anomaly:2009,OutlierSurvey}. In Section~\ref{sec:approach:prior}, we introduce two approaches (SKM and SHAC) that rely on (c).}

%  One-Class Support Vector Machines (OCSVM) \cite{scholkopf2001estimating}, Cluster-Based Local Outlier Factor (CBLOF) \cite{duan2009cluster}, LUNAR \cite{goodge2022lunar}
% OCSVM, CBLOF, COF, PCA and KPCA.

% The objective of anomaly detection techniques is to distinguish abnormal patterns or outliers from the majority of data points. Anomaly detection techniques aim to capture these deviations and provide an early warning system. 
Below, we describe the state-of-the-art anomaly detection approaches that we selected for \APPR.

%approaches applicable to our context (numeric datasets).
%There are numerous clustering based anomaly detection techniques such as Local Outlier Detection and Isolation Forest, moreover other techniques such as k-means, DBSCAN and Agglomerative clustering can also be used indirect for anomaly detection based on distance. Although anomaly detection is a critical task, but the help of domain expert, aforementioned techniques can be used to perform clustering with objective to cluster anomalies and normal data points in different clusters. 

% IF
\emph{IF} is a powerful method for anomaly detection in high-dimensional datasets. It is based on the intuition that anomalies are often isolated from the majority of data points and can, therefore, be identified quicker. The IF algorithm constructs a binary tree-like structure known as an \emph{Isolation Tree} by randomly selecting features and splitting the data at random values within those features. This partitioning process is repeated recursively until each data point is isolated within a leaf or all the data points in a leaf have the same values. Anomalies are then identified as those data points that require fewer splits to isolate, making IF particularly effective at detecting outliers or anomalies within complex and high-dimensional datasets. This algorithm has gained popularity for its speed and accuracy in anomaly detection tasks, making it a valuable tool in various domains, including fraud detection, network security, and quality control.

K-Nearest Neighbor (KNN) is a machine learning algorithm used for both classification and anomaly detection; it is the building block for LOF. In KNN, the \emph{K} represents the number of nearest neighbors to consider when making a prediction for a new data point. The fundamental idea behind KNN is that data points are similar to those in their proximity. 
%For classification tasks, KNN determines the class of a new data point by examining the majority class among its K-nearest neighbors. Each neighbor's class contributes to the prediction with equal weight, and the class with the highest representation among the neighbors is assigned to the new data point. 
The application of KNN principles to anomaly detection led to different approaches. One approach, for example, consists of identifying as anomalous those data points with a K$^{\mathit{th}}$ neighbour having a distance above a threshold \emph{d}~\cite{KnorrKNN}; another approach identifies as anomalous the N points having the largest distance from the K$^{\mathit{th}}$ neighbour~\cite{RamaswamyKNN}. However, the former approach requires the identification of an ideal threshold, which might be difficult to achieve; the latter, does not take into account the local density of the neighbourhood (e.g., K$^{\mathit{th}}$ neighbour might be far away but the others not).
% LOL or LOF
LOF, instead, 
%developed by Breunig et al. in 2000, is a technique for detecting local anomalies or outliers within datasets. 
%Unlike traditional global outlier detection methods, 
adopts a density-based approach, considering the local neighborhood density of each data point. 
It leverages the concept of KNN to assess local density but, different from the approaches described above,
computes an anomaly score as the average of the distances of a point from its K nearest neighbours. The anomaly score computed by the LOF
 algorithm 
 %measure known as the Local Outlier Factor (LOF) 
 quantifies how isolated a point is within its immediate surroundings. 
Low anomaly score values indicate that a data point is similar in density to its neighbors, while high LOF values signify that a point belongs to a less dense neighbourhood and, likely, is an outlier. In its default configuration, LOF reports as anomalous those data points with an anomaly score above 1.5. 
%By comparing these distances, LOF effectively identifies outliers that exhibit distinctive local density patterns. 
%LOF demonstrated being an effective anomaly detection technique, finding applications in various domains, including fraud detection, network security, and quality control, particularly when anomalies exhibit varying local densities within the data.

\MAJOR{1.11, 5.4}{Future work includes extending \APPR with additional algorithms such as 
One-Class Support Vector Machines (OCSVM) \cite{scholkopf2001estimating}, or Cluster-Based Local Outlier Factor (CBLOF) \cite{duan2009cluster}. 
%and LUNAR~\cite{goodge2022lunar}, which relies on graph neural networks. 
Additionally, semi-supervised algorithms might be considered to take advantage of long-term deployments or feedback-based detection~\cite{OutlierSurvey}.}

% !TEX root =  MAIN.tex

\section{Proposed Approach: \APPR}
\label{sec:approach}

FISTS relies on the intuition that security vulnerabilities induced by an SDN configuration change (hereafter, \emph{SDN reconfiguration}) might be detected by identifying port state changes that look anomalous if compared to the other port state changes. \APPR implements a field-based testing approach that probes the network before and after an SDN reconfiguration to determine port states, and then, after identifying port state changes, even in the presence of re-assigned IP and MAC addresses, executes anomaly detection algorithms to determine the vulnerable changes that affect specific hosts. 

%Specifically operates to identify and capture the change in network topology by comparing the initial and upgraded configuration of the network. The change in network configuration is interpreted by host matching module which saves the difference of both configurations in a representative way in \textit{.csv} format for machine learning models.

\begin{figure*}[ht] 
\centering
	\includegraphics[width=16cm]{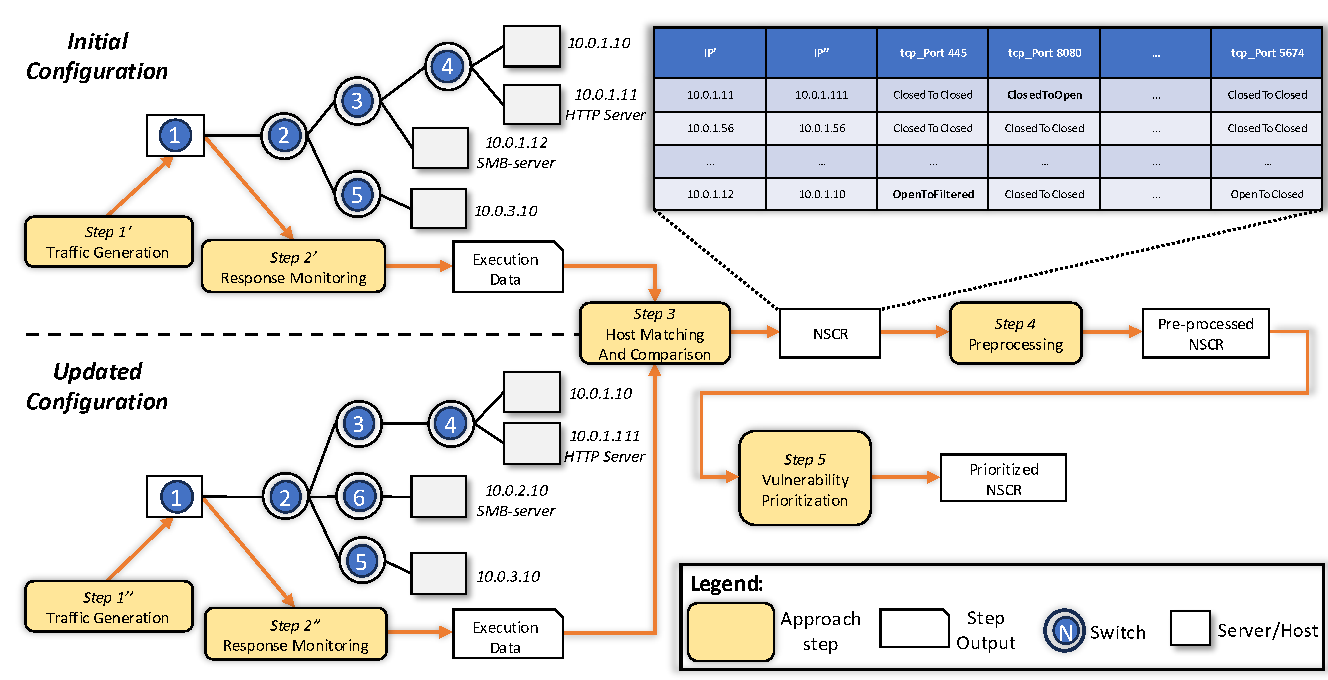}
%  Rounded yellow rectangle: approach step 
% white rectangles: step output
% gray rectangle: server/host
 \caption{Overview of \APPR. It shows \APPR steps along with example outputs. The left side shows a network being tested by \APPR before and after reconfiguration. The NSCR output is exemplified by the provided table.}
  \label{fig:approach:topology}
\end{figure*}

\APPR can detect reconfigurations leading to security vulnerabilities due to erroneous port states including the one exemplified in Fig.~\ref{fig:approach:topology} (see Section~\ref{sec:empirical:subjects} for other cases). The original SDN configuration shows a Server Message Block (SMB) server (operates on port 445) connected to switch 3 and a high-priority flow rule on switch 2 that, to increase confidentiality by reaching only the subnet with the SMB server, forwards all traffic for port 445 to switch 3. In an update, the engineers move the SMB server to a dedicated switch (i.e., switch 6) but forget to update the rule on switch 2 to forward SMB traffic to the new destination switch 6 instead of switch 3. Consequently, all the traffic for the SMB server is routed to the wrong subnet, which poses serious confidentiality issues even if packets may be re-routed to reach switch 6 (not shown in the picture). \MAJOR{4.2}{Note that Fig.~\ref{fig:approach:topology} assumes proactive SDN flows (i.e., packets flow is preconfigured);
although reactive SDN flows (i.e., rules are created as packets come into the switch) may prevent such mistakes, they cannot be used in several context (e.g., because of confidentiality requirements preventing the free routing of packets).}

%SMB server uses \textit{tcp} protocol hence no acknowledgment (ACK) for packets is sent to the sender.

\APPR works in four steps, depicted in Fig.\ref{fig:approach:topology} and described below. Note that both Step 1 and Step 2 are performed twice, before and after an SDN reconfiguration.

\subsection{\APPR Steps 1 and 2: Traffic Generation and Response Monitoring}

In Step 1, \APPR generates network traffic for reconnaissance purposes, the responses generated by the hosts in the network are collected in Step 2 and processed to produce information about the port states of each host in the network. Our current implementation of \APPR relies on NMAP for both Step 1 and Step 2 because NMAP can generate data for network reconnaissance (i.e., \APPR Step 1) and produces
scan result files in XML format (i.e., \APPR Step 2).
We rely on NMAP TCP SYN and UDP scans.
However, other tools might be integrated in the future to extend our capabilities; for example, we may rely on FragScapy~\cite{Fragscapy} to collect additional information masked by stateful firewalls. At a high-level, the result of Step 2 is a dataset where each data entry provides the following information for one host in the network: 
\begin{itemize}
    \item MAC address
    \item IP address
    \item for all the ports,
    \begin{itemize}
    \item whether they are UDP or TCP
    \item if they are in one of the following states: Open, Closed, Filtered, Open$|$Filtered
    \end{itemize}
\end{itemize}

%For scanning the network we use Nmap which is an open-source network reconnaissance utility for scanning devices in the network and services hosted on them. In Fig.\ref{fig:approach:topology} we use nmap to generate traffic(for scanning) from host 1 for network reconnaissance and save the scan result in \textit{.xml} file. We do the same after the update in the configuration and save the result in another \textit{.xml} file.

\begin{table*}[]
\caption{Example execution data collected by NMAP for an initial configuration (O: open, C: closed)}
\label{initial_config_table}
\centering
\scriptsize

\begin{tabular}{cccccccccc}
\hline 
\textbf{IP} & \textbf{MAC}  & \textbf{tcp\_Port 22} & \textbf{udp\_Port 22} & \textbf{tcp\_Port 123} & \textbf{udp\_Port 123} & \textbf{tcp\_Port 445} & \textbf{...}          & \textbf{tcp\_Port 8080} & \textbf{udp\_Port 8080} \\ \hline
10.0.1.56 & 00:1A:2B:3C:4D:5E & C                     & C                     & C                      & \textbf{O}             & C                      & ...                   & C                       & C                       \\ \hline
10.0.1.12 & 08:15:23:42:67:94 & C                     & C                     & C                      & C                      & \textbf{O}             & \textit{\textbf{...}} & C                       & C                       \\ \hline
10.0.1.11 & 2A:9F:C7:0E:3D:81 & C                     & C                     & C                      & C                      & C                      & ...                   & \textbf{O}                       & C              \\ \hline
10.0.3.10 & 5C:73:2F:A1:6E:B0 & \textbf{O}            & C                     & C                      & C                      & C                      & ...                   & C                       & C                       \\ \hline

\hline

\end{tabular}
\end{table*}

\begin{table*}[]
\caption{Example execution data collected by NMAP for an updated configuration}
\label{Updated_configuration_table}
\centering
\scriptsize
\begin{tabular}{cccccccccc}
\hline 
\textbf{IP} & \textbf{MAC} & \textbf{tcp\_Port 22} & \textbf{udp\_Port 22} & \textbf{tcp\_Port 123} & \textbf{udp\_Port 123} & \textbf{tcp\_Port 445} & \textbf{...}          & \textbf{tcp\_Port 8080} & \textbf{udp\_Port 8080} \\ \hline
10.0.1.56 & 00:1A:2B:3C:4D:5E  & C                     & C                     & C                      & \textbf{C}             & C                      & ...                   & C                       & C                       \\ \hline
10.0.2.10 & 08:15:23:42:67:94 & C                     & C                     & C                      & C                      & \textbf{F}             & \textit{\textbf{...}} & C                       & C                       \\ \hline
10.0.1.111 & 2A:9F:C7:0E:3D:81 & C                     & C                     & C                      & C                      & C                      & ...                   & \textbf{O}                       & C              \\ \hline
10.0.3.10 & 5C:73:2F:A1:6E:B0 & \textbf{O}            & C                     & C                      & C                      & C                      & ...                   & C                       & C                       \\ \hline

\hline

\end{tabular}
\end{table*}

Tables~\ref{initial_config_table} and~\ref{Updated_configuration_table} provide examples of data entries collected for the two configurations appearing in Fig.~\ref{fig:approach:topology}. To enable anomaly detection, features should be consistent across entries; hence, for every port, we capture information for both the TCP and the UDP protocol as separate features (i.e., columns).

%consequently, if a port is open for TCP then it will be closed for UDP.

%It is noticeable that, in the initial configuration, the SMB server has IP XX, MAC address XX, and ports ... being open. In the updated configuration, the SMB server has IP YY, MAC address XX, and ports ... being open.
%In the scan of an updated configuration, packets from Nmap are unable to reach SMB-server therefore Nmap reports port 445 this happens due the the presence of high-priority rule at switch 2 which sends all the traffic for port 445 to switch 3.

\subsection{\APPR Step 3: Host Matching and Comparison}

%Scans files(.xmls) for initial and updated configuration are processed by 
In Step 3, the Host Matching and Comparison (HMC) module processes the data collected before (by Step 2') and after (by Step 2") a reconfiguration to (1) determine what data entries belong to a same host in the two configurations and (2) determine what are the port state changes for each host. 

Determining what data entries belong to a same host is necessary because SDN reconfigurations may change the IP or MAC address of one host (e.g., a server) without changing anything else. Not noticing hosts with addresses being changed may lead to incorrect identification of port state changes. For example, in Fig.~\ref{fig:approach:topology}, the IP of the SMB server may change from 10.0.1.12 to 10.0.2.10 (MAC address remains the same); without noticing that the SMB sever has been simply moved to another IP address, \APPR may interpret all the open ports (e.g., 445) of IP 10.0.1.12 as becoming closed and port 445 of IP 10.0.2.10 as becoming filtered, which may misguide the \APPR anomaly detection algorithm used in Step 5 or the end-user processing \APPR results. Similarly, a host with a Web server has been moved from IP 10.0.1.11 to IP 10.0.1.111, 
without noticing that the host is likely the same, we may interpret all the ports of IP 10.0.1.11 as becoming closed and port 8080 of IP 10.0.1.111 as becoming open, which may lead to false positives (e.g., such bulk changes of port states may be reported as anomalous by an anomaly detection algorithm).

% !TEX root =  ../Main.tex

%\vspace{-3mm}
\begin{figure}[tb]

\begin{algorithmic}[1]

%\footnotesize
\scriptsize
\Require $\mathit{DS}_{\mathit{initial}}$, data set collected from the initial SDN configuration
\Require $\mathit{DS}_{\mathit{updated}}$, data set collected from the updated SDN configuration
%\Require \emph{dataProvider}, an object that exposes the data collected by the crawlers
\Ensure $\mathit{MATCH}$, a table whose rows contain the values of two matching entries from $\mathit{DS}_{\mathit{initial}}$ and $\mathit{DS}_{\mathit{updated}}$, plus a match score

\Comment{Compute a matching score for all the possible pairs of entries}

\State $\mathit{scored} \gets$ empty list
\For {$e_n$ in initial configuration set} \label{algo:match:N}
    \For {$e_m$ in updated configuration set} \label{algo:match:M}
    \If{IP address of $e_n$ and $e_m$ match} \label{algo:match:IPmatch}
        \State $\mathit{score\_ip} \gets 0.2$
    \EndIf
    \If{MAC address of $e_n$ and $e_m$ match}
        \State $\mathit{score\_mac} \gets 0.2$
    \EndIf
    
    \Comment{Compute $\mathit{score\_ports}$}
    
    \State $\mathit{matches} \gets 0$
    \State $\mathit{ports} \gets 0$
    \For {any port $p$ being not closed in either $e_n$ and $e_m$}
        \State $\mathit{ports} \gets \mathit{ports} + 1$
        \If{state of port $p$ in $e_n$ and $e_m$ match}
        \State $\mathit{matches} \gets \mathit{matches} + 1$
    \EndIf
    \EndFor\label{algo:match:portEnd}
    \State $\mathit{score\_ports} \gets \frac{\mathit{matches}}{\mathit{ports}} * 0.6$

    \Comment{Score of the entry pair}
    \State $\mathit{matching\_score} \gets score\_ip + score\_mac + score\_port$
    \State $\mathit{scored} \gets \mathit{scored} \cup \langle e_n.\mathit{IP}, e_m.\mathit{IP}, \mathit{matching\_score}\rangle$
\EndFor
\EndFor
\State sort $\mathit{scored}$ based on $score$ \label{algo:match:sort}
\State $\mathit{MATCH} \gets$ empty set
\For {$\langle e_n.\mathit{IP}, e_m.\mathit{IP}, \mathit{matching\_score}\rangle$ in $scored$} \label{algo:match:storeStart}
    
    \Comment{Check if $e_n$ is not already matching another entry}
    
    \If { $e_n.\mathit{IP}$ in $\mathit{assigned\_initial}$ }
    \State \textbf{continue}
    \EndIf
    
    \Comment{Check if $e_m$ is not already matching another entry}
    
    \If {  $e_m.\mathit{IP}$ in $\mathit{assigned\_updated}$}
    \State \textbf{continue}
    \EndIf
    
    \Comment{Add the matching pair to $\mathit{MATCH}$}
    
    \State $\mathit{MATCH} \gets \mathit{MATCH} \cup \langle e_n.\mathit{IP}, e_m.\mathit{IP}, \mathit{matching\_score}\rangle$

    \Comment{Update the list of processed IPs}

    \State $\mathit{assigned\_initial} \gets \mathit{assigned\_initial} \cup e_n.\mathit{IP}$
    \State $\mathit{assigned\_updated} \gets \mathit{assigned\_updated} \cup e_m.\mathit{IP}$

\EndFor    \label{algo:match:storeEnd}

\For {$e_n$ in initial configuration set} \label{algo:match:N}\label{algo:unmatched:start}
\If { $e_n.\mathit{IP}$ not in $\mathit{assigned\_initial}$ }
    \State $\mathit{MATCH} \gets \mathit{MATCH} \cup \langle e_n.\mathit{IP}, null\_IP, 1.0\rangle$
\EndIf
\EndFor
\For {$e_m$ in updated configuration set} \label{algo:match:N}
\If { $e_m.\mathit{IP}$ not in $\mathit{assigned\_updated}$ }
    \State $\mathit{MATCH} \gets \mathit{MATCH} \cup \langle null\_IP, e_n.\mathit{IP}, 1.0\rangle$
\EndIf
\EndFor\label{algo:unmatched:end}
\end{algorithmic}

\caption{\APPR's host matching algorithm.}
%\vspace{-0.2cm}
\label{algo:matching}
\end{figure}

%which compares hosts in both scans to determine the respective matching host based on IP address, MAC address, and state of the ports to assign a relativity score between 0$~$1. 
Since hosts with a change in their IP or MAC address may also present changes in their port states (e.g., the SMB server in our running example); it is not possible to simply look at matching port states to identify the hosts assigned to a different IP or MAC address. Therefore, we have developed a dedicated greedy algorithm; its pseudo-code appears in Fig.~\ref{algo:matching}.
Our matching algorithm performs $n \times m$ comparisons between the $n$ data entries from the initial configuration and the $m$ data entries from the updated configuration (Lines~\ref{algo:match:N} and~\ref{algo:match:M} in Fig.~\ref{algo:matching}). For each data entry pair, it 
determines if the IP, MAC address, and state of each port match (Lines~\ref{algo:match:IPmatch} to~\ref{algo:match:portEnd}).
Such information is used to assign a matching score to each pair, as follows:
\vspace{-1em}
$$matching\_score = score\_ip + score\_mac + score\_ports$$
The value of $score\_ip$ is set to $0.2$ when two IPs match, otherwise it is set to $0$.
The value of $score\_mac$ is set to $0.2$ when MAC addresses match, otherwise it is set to $0$.
The value of 
$score\_ports$ is computed by considering all the ports reported as not closed in at least one the two entries, and by multiplying the proportion of matching port states by $0.6$.
The algorithm then sorts all the pairs based on their matching score (Line~\ref{algo:match:sort}) and stores the matched pairs till all the IPs in the two datasets had been considered (Lines~\ref{algo:match:storeStart} to~\ref{algo:match:storeEnd}). Our choice for the values assigned to $score\_ip$, $score\_mac$, and $score\_ports$ enables ensuring that a host that, in the updated configuration, changes IP or MAC and a limited subset of its ports is still matched with the correct entry in the initial configuration. Unmatched IPs correspond to hosts being added or removed from the network and are thus matched with a generated entry having null IP (Lines~\ref{algo:unmatched:start} to~\ref{algo:unmatched:end}).

% You: What happens if an IP was not present in the initial configuration? Is it still compared to other IPs or what?

% What happens if the initial configuration has more IPs than the updated one? 
% Jan 9, 2024 4:10 PM • 
%  • 

% Jahanzaib MALIK: in that case, IP and MAC is none, and all ports go from close to open(incase if any port is open/filtered)

% If all ports in the new IP are closed, than IP and MAC stays none but port go from closed to closed

\begin{table}[]
\caption{Example of NSCR state changes}
\label{NSCR_state_table}
\centering
\scriptsize
\begin{tabular}{ccc}
\hline
\textbf{Initial} & \textbf{Updated} & \textbf{Label} \\ \hline\hline
Closed                         & Open                           & ClosedToOpen              \\ \hline
Open                           & Closed                         & OpenToClosed              \\ \hline
Open$|$Filtered                         & Closed                         & OpenFilteredToClosed              \\ \hline 
\end{tabular}
\end{table}

Our matching algorithm leads to pairs of IPs, each modeling a single host. 
%which may have changed its IP or MAC address after the SDN reconfiguration. 
They are used by \APPR to construct, using the execution data collected by Step 3, a table, called \emph{Network State Change Report} (NSCR), that reports, for each host, its port state changes. In the NSCR, each port state change is captured by a label that joins the name of the port state in the initial and the updated configuration for the selected port; Table~\ref{NSCR_state_table} shows a few examples of how port state change labels are generated from specific port states in the initial and updated configuration. In total, we have 16 different label (i.e., 4 labels for the initial state times 4 labels for the updated state).
An example NSCR is shown in Fig.\ref{fig:approach:topology}, each column contains information about the state change observed for one port working with a specific protocol (UDP or TCP). \MAJOR{2.3}{Please note that we also track ports that do not change their state (e.g., \emph{tcp\_Port 445} in the first row of Fig.\ref{fig:approach:topology}'s table, labeled as \emph{ClosedToClosed}).}
%The relativity score is further balanced between IP address, MAC address and state of the ports based on weight that is 0.2, 0.2, and 0.6 respectively. 
%Moreover, HMC also creates a table of unique hosts in both scans to create a dataset with features i.e., IP\_V1, IP\_V2, MAC\_V1, MAC\_V2, Ports (1$~$n), IP\_Changed, MAC\_Changes. 
%Description of features are presented in Table \ref{Table:data:features}.
%The representation of the dataset can be seen in the Fig.\ref{fig:approach:topology} with ports and their states. The table is exported as \textit{.csv} which is further used by machine learning algorithms for vulnerability prioritization.

\MAJOR{4.3}{In the presence of  firewalls, port scanning with NMAP will lead to most or all ports being shown as filtered (see Table~\ref{table:NMAP:portstates}). However, this behavior does not compromise the applicability of \APPR. Indeed, service providers testing their own infrastructure can configure scanning nodes appropriately, if needed. For example, in the presence of firewalls that prevent responses from subnets, if the scanning node is on the WAN,
%; 
NMAP will report that all the subnet nodes have filtered ports; however, if after a configuration change, a subnet provides access from WAN to an SMB server on 8080, \APPR will report \emph{FilteredToOpen}, which is a state change (note that without a firewall  we would observe \emph{ClosedToOpen}), which may still enable anomaly detection. If access to some services of a given private subnet is provided only to other private subnets, multiple scanning nodes might be used (e.g., one per subnet). Further, approaches relying on packet fragmentation such as hping~\cite{hping}, FragScapy~\cite{Fragscapy}, and some NMAP extensions~\cite{nmapFirewall} enable port scanning in the presence of firewalls and can be integrated into \APPR.}

\subsection{\APPR Step 4: Preprocessing}
\label{sec:step:four}

In Step 4, the NSCR is preprocessed to enable the application of anomaly detection algorithms. Indeed, such algorithms usually work with numerical data while NSCR entries contain textual categorical data. To transform our data, we cannot apply integer encoding (i.e., replace each unique category label with an integer) because it would introduce an arbitrary ordering across labels (e.g., 'OpenToClosed' being closer to 'OpenToFiltered' than to 'ClosedToOpen)', which is a bad practice~\cite{burkov2020machine}; therefore, we apply one-hot encoding. Since one-hot encoding replaces each column with a number of boolean columns matching the number of distinct possible values, we derive 16 columns for each TCP and UDP port column. 
%An example is shown in Table~\ref{NSCR_state_table_Detailed}.
%\input{tables/NSCR_state_table_Detailed}

In Step 4, \APPR may also perform an additional, optional, task, which consists of \emph{pruning} entries that do not present any change in their state. Our intuition is that reconfigurations not introducing any change in the port states of a host unlikely introduce a vulnerability affecting such host because the set of open, filtered, and closed ports remains the same, and thus the host works as in the previous configuration. Instead, keeping such entries may increase the likelihood of false positives; for example, in the presence of several hosts without port state changes, Isolation Forest may create several trees where all the sampled data entries match, have a minimal distance from the root, and are thus erroneously reported as anomalous. 

\subsection{\APPR Step 5: Vulnerability Prioritization}
\label{sec:approach:prior}

In Step 5, \APPR sort all the entries in the NSCR according to their likelihood of being affected by a vulnerable configuration, such likelihood is captured by an anomaly score generated by an anomaly detection algorithm. An example output is shown in Table~\ref{NSCR_state_table_Detailedddd}.

% \begin{table*}[]
% \caption{NSCR alongside anomaly score}
% \label{NSCR_AnomalyScore}

% \centering
% \begin{tabular}{ccccccc}
% \hline \hline
% \textbf{IP'} &\textbf{IP"} & \textbf{Initial Config} & \textbf{Updated Config} & \textbf{...} & \textbf{Label} & \textbf{Anomaly Score} \\ \hline \hline
% 192.168.56.113 & 192.168.56.113 &  Closed                         & Open          & ...                 & 1              & 0.8                    \\ \hline
% 192.168.56.114 & 192.168.56.214 & Open                           & Closed            & ...             & 1              & 0.74                   \\ \hline
% 192.168.56.115 & 192.168.56.115 & Open                           & Filtered           & ...            & 1              & 0.53                   \\ \hline
% 192.168.56.116 & 192.168.56.116 & Open                           & Open              & ...             & 0              & 0.4                    \\ \hline
% 192.168.56.117 & 192.168.56.117 & Closed                         & Closed           & ...              & 0              & 0.3                    \\ \hline \hline
% \end{tabular}
% \end{table*}

\begin{table*}[]

\caption{Prioritized NCSR report example}
\label{NSCR_state_table_Detailedddd}
\centering
\scriptsize
\begin{tabular}{cccccccccc}
\hline 
\textbf{IP'}   & \textbf{IP"}   & \textbf{tcp\_Port 22}     & \textbf{udp\_Port 22}           & \textbf{tcp\_Port 445} &  \textbf{udp\_Port 445}              & \textbf{tcp\_Port 8080}                  & \textbf{...}                            & \textbf{udp\_Port 5674 }   & \textbf{Anomaly Score}     \\ \hline 
10.0.1.12 & 10.0.2.10 & ClosedToClosed & ClosedToClosed  & \textbf{OpenToFiltered}        & ClosedToClosed           & ClosedToClosed & ...                 & ClosedToClosed & 0.97 \\ \hline 

10.0.1.22 & 10.0.1.22 & ClosedToClosed & ClosedToClosed  & ClosedToClosed                   & ClosedToClosed& ClosedToClosed & ...                 & \textbf{OpenToFiltered}  & 0.93 \\ \hline
10.0.1.56 & 10.0.1.56 & ClosedToClosed & ClosedToClosed  & ClosedToClosed                   & ClosedToClosed & ClosedToClosed& ... & ClosedToClosed & 0.82 \\ \hline
10.0.3.10  & 10.0.3.10  & ClosedToClosed & ClosedToClosed  & ClosedToClosed & ClosedToClosed& ClosedToClosed & ...                 & ClosedToClosed & 0.77\\ \hline

... & ... & ...&  ...                  & ... & ...         &...        & ...& ...  & ...\\
\hline
10.0.1.11 & 10.0.1.111 & ClosedToClosed & ClosedToClosed  & ClosedToClosed      & ClosedToClosed           & \textbf{OpenToOpen} & ...                 & ClosedToClosed & 0.39 \\ \hline

\end{tabular}
\end{table*}

% \begin{tabular}{ccccccccc}
% \hline \hline
% \textbf{IP'}   & \textbf{IP"}   & \textbf{Port 22}        & \textbf{Port 80}        &  \textbf{Port 445}                         & \textbf{Port 8080}      & \textbf{...}                            & \textbf{Label } & \textbf{Anomaly Score }     \\ \hline \hline
% 192.168.56.113 & 192.168.56.113 &  ClosedToClosed & ClosedToClosed & ClosedToClosed                   & ClosedToClosed & ...                 & 1& 0.96 \\ \hline
% 192.168.56.114 & 192.168.56.214 &  ClosedToClosed & ClosedToClosed & ClosedToClosed                   & \textit{\textbf{OpenToClosed}} & ... & 1& 0.72 \\ \hline
% 192.168.56.115  & 192.168.56.115  & ClosedToClosed & ClosedToClosed  & ClosedToClosed & ClosedToClosed & ...                 & 1 & 0.65 \\ \hline
% ... & ... & ...&  ... & ...                 & ... & ...                 & ...& ... \\
% \hline
% 192.168.56.118 & 192.168.56.118 & ClosedToClosed &  ClosedToClosed & \textit{\textbf{OpenToFiltered}}                   & ClosedToClosed & ...                 & 0& 0.3 

% \\ \hline \hline

% \\\\
% \end{tabular}

Our key intuition is that the prioritization of all the hosts enables the definition of a human-in-the-loop process that overcomes the limitations of existing anomaly detection approaches relying on predefined thresholds.
Indeed, existing anomaly detection approaches select a subset of data entries as anomalous based on some arbitrary thresholds, which are defined either for the distance between dataset features~\cite{RamaswamyKNN} or for the generated anomaly scores~\cite{isolation_forest,LOF}. However, since features may change from context to context (e.g., different SATCOM providers may change different sets of ports in different SDN reconfigurations), thresholds that are successful in one case study may not work with others. Instead, it would be better to enable the end-user stop inspecting anomalous hosts when there is evidence that what reported by \APPR does not help discovering vulnerabilities; we achieve such objective by using as stopping condition (SC) the number of consecutive false positives (CFP) observed, if $\mathit{FP} \ge \mathit{SC}$ the engineer can stop inspecting hosts. 

By relying on the number of consecutive false positives as a stopping condition, unlike related work, we do not assume that anomalies might be identified with the same distance thresholds in different case study subjects, but we look for evidence that hosts with an anomaly score below a certain value aren't vulnerable, which is achieved by verifying that there are SC consecutive false positives. In our empirical evaluation, we assess the performance of different configurations of \APPR, with different SCs.

We call the approach above \emph{\APPR hosts selection} and it has the objective of reducing SDN maintenance costs by enabling engineers to inspect only a subset of the potentially vulnerable hosts. However, companies delivering critical services may require their engineers to inspect all the hosts, to ensure the absence of any vulnerability. In such case, engineers should inspect all the hosts in the order provided by \APPR, we call such approach \emph{hosts prioritization}. When applied for hosts prioritization, \APPR should enable engineers identify all the vulnerable hosts at the beginning of their investigation, thus minimizing the time required to discover and fix vulnerabilities.

To sort NSCR hosts, \APPR should rely on anomaly detection algorithms producing an anomaly score that can be used for prioritization. Unfortunately, well-known algorithms for anomaly detection (i.e., IF and LOF) select a subset of dataset entries as anomalous instead of sorting dataset entries according to their likelihood of being anomalous. However, internally, they rely on an anomaly score to sort dataset entries and then select the ones with the highest anomaly score; therefore, we defined two approaches that rely on such internal information. We call \emph{Sorted Isolation Forest (SIF)} our approach relying on the anomaly score generated by the IF algorithm. We call \emph{Sorted KNN (SKNN)}, the algorithm that sorts dataset entries based on the anomaly score computed by LOF, which consists of the average pairwise distance computed using a KNN approach. 

Although effective, IF and LOF aren't the only approaches capable of extracting anomaly information from a dataset. Precisely, since some vulnerabilities may present similar effects (i.e., they lead to the same port changes), we believe that clustering algorithms can detect hosts with similar changes, and thus enable the detection of hosts with similar vulnerabilities. However, clustering algorithms do not sort data entries; but vulnerable configurations tend to affect a small subset of the hosts (e.g., because they are introduced by corner cases), and thus we can prioritize the inspection of hosts based on the size of the cluster they belong to. Precisely, clusters with few items are more likely due to a vulnerable configuration and should thus be inspected first. Within a cluster, entries are inspected in a random order.

To perform clustering, we rely on K-means and HAC (see Section~\ref{sec:background}). We selected K-means because it is a standard baseline clustering approach. We selected HAC because our preprocessed dataset results from the application of one-hot encoding, and thus the distance between two data entries would capture the number of port state changes that do not match between two entries; in other words, hosts with the same port state changes will be very close. Since we assume that a vulnerable configuration affects hosts in a same way (i.e., port states change similarly), a hierarchical clustering algorithm that builds clusters by joining hosts that are close to each other, seems an appropriate fit for the identification of clusters with vulnerable hosts. We refer to our strategy to prioritize hosts based on clustering algorithms as \emph{sorted K-means} (SKM) and \emph{sorted HAC} (SHAC).

\subsection{\APPR usage}

The \APPR approach is implemented as a pipeline;
a \emph{\APPR pipeline consists} of all the components automating Steps 1 to 5. A pipeline configuration is set by (1) enabling/disabling the pruning component, (2) selecting the anomaly detection algorithm to be used (i.e., SIF, SKNN, SKM, SHAC), and, if needed, setting its hyper-parameters (i.e., K for SKNN) (3) selecting the stopping condition (i.e., the number of false positives after which engineers should stop inspecting results, if engineers aim to apply \APPR for \emph{hosts selection}).

\section{Empirical Evaluation}
\label{sec:empirical} 

We performed an empirical evaluation that aims at addressing the following research questions (RQs):

\newcommand{\RQone}{Does pruning improve \APPR results?}
\emph{RQ1.} \emph{\RQone}
The pruning tasks in \APPR preprocessing step (i.e., ignoring hosts not affected by any state change) may influence the execution of anomaly detection algorithms. For example, without pruning, a density-based algorithm like LOF or SKNN may end-up computing anomaly scores that are very similar for non-vulnerable hosts with no state changes and for hosts with one (vulnerable) state change (same density), thus leading to prioritized hosts where several false positives may appear before a true positive. Further, as discussed in Section~\ref{sec:step:four}, the absence of pruning may affect IF as well.
%sampling-based approach like IF (e.g., without preprocessing, hosts with a few ports having a state change may be discarded as not anomalous in a sampling based approach like IF). 
Therefore, we aim to determine if, overall, the best results are observed with or without pruning.
 
% What \APPR configuration performs better for a varying number of vulnerable hosts? Since we cannot make assumptions on the number of vulnerabilities affecting the system under test, we aim to perform an ablation study to determine the anomaly detection component and the false positive threshold performing better for varying proportion of vulnerabilities.

\newcommand{\RQtwo}{What \APPR pipeline leads to the best results?}
\emph{RQ2.} \emph{\RQtwo{}}
The performance of a \APPR pipeline may depend on the characteristics of the SDN under test; for example, the total number of hosts with valid and invalid state changes and the number of ports with valid/invalid state changes may lead to different performance results for the \APPR anomaly detection algorithms and, consequently, may affect what would be the best stopping condition. We aim to assess what is the \APPR pipeline leading to the best results with different datasets.
 
%configuration performs better in realistic scenarios? This study aims to confirm RQ1 results using historical data concerning our industry partner.
\newcommand{\RQthree}{How does the host matching component impact on \APPR results?}
\emph{RQ3.} \emph{\RQthree{}} We aim to assess the accuracy of the host matching component and its impact on \APPR results (e.g., the proportion of false alarms due to inaccurate host matching).
%Potentially we can assess how mistakes influence resukts

\newcommand{\RQfour}{How does \APPR scale?}
\emph{RQ4.} \emph{\RQfour} We aim to report on the scalability of the approach for a network having the characteristics (e.g., configured ports) of our industry partner's, which we expect to be representative of SATCOM organizations. We aim to assess both network probing and the anomaly detection component.

\subsection{Experiment Setup}
\label{sec:empirical:subjects}

We target SATCOM providers that rely on SDNs for network communications, like our industry partner \SES{}~\cite{SES}. To perform experiments that reflect production conditions and, at the same time, share our datasets without breaking confidentiality agreements, we considered public data about \SES{} networks. \MAJOR{4.5}{Although SES cannot disclose the nature (e.g., SDN or traditional) of the networks monitored by Shodan, SES engineers been confirmed that they share similar characteristics (number of nodes, open ports) with SES-managed SDN networks.}

\begin{figure*}[ht]
    \centering
    \begin{subfigure}{0.36\textwidth}
        \centering
        \includegraphics[width=\textwidth]{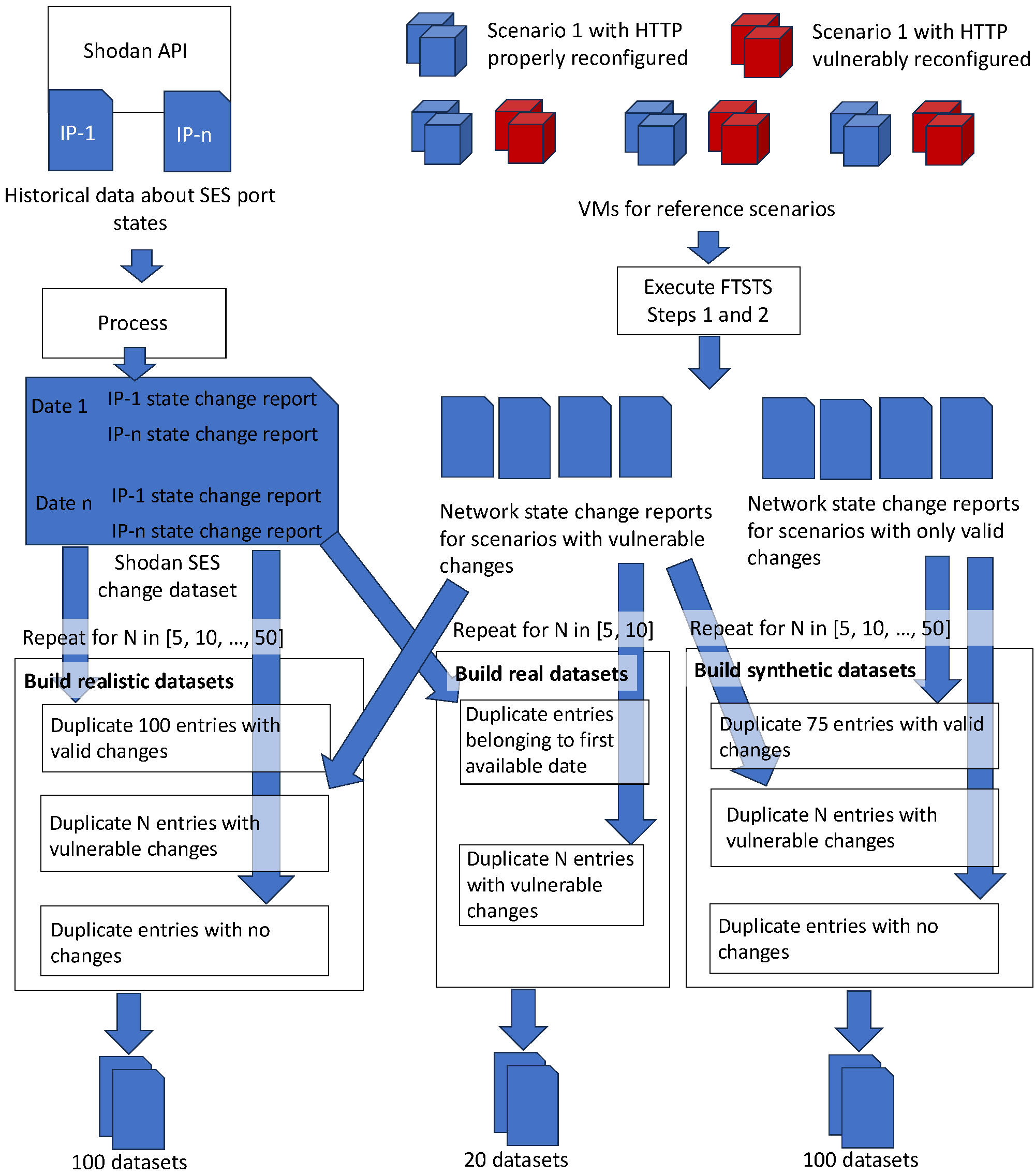}
        \caption{Datasets creation procedure}
        \label{fig:datasets:creation}
    \end{subfigure}
    \hfill % adds horizontal space between figures
    \begin{subfigure}{0.30\textwidth}
        \centering
        \includegraphics[width=\textwidth]{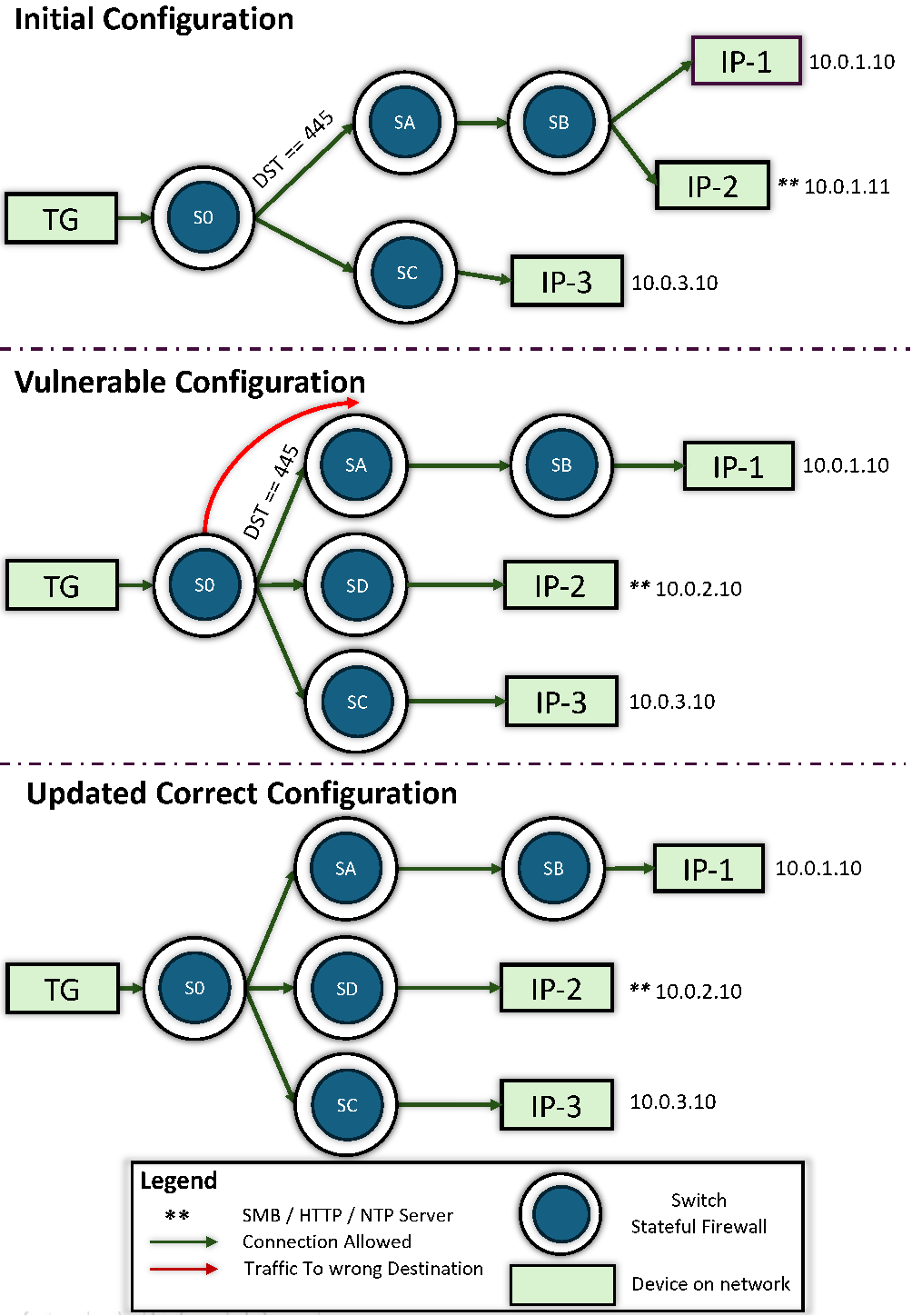}
        \caption{Scenario with redundant rule}
        \label{fig:scenario:redundant}
    \end{subfigure}
    \hfill
    \begin{subfigure}{0.25\textwidth}
        \centering
        \includegraphics[width=\textwidth]{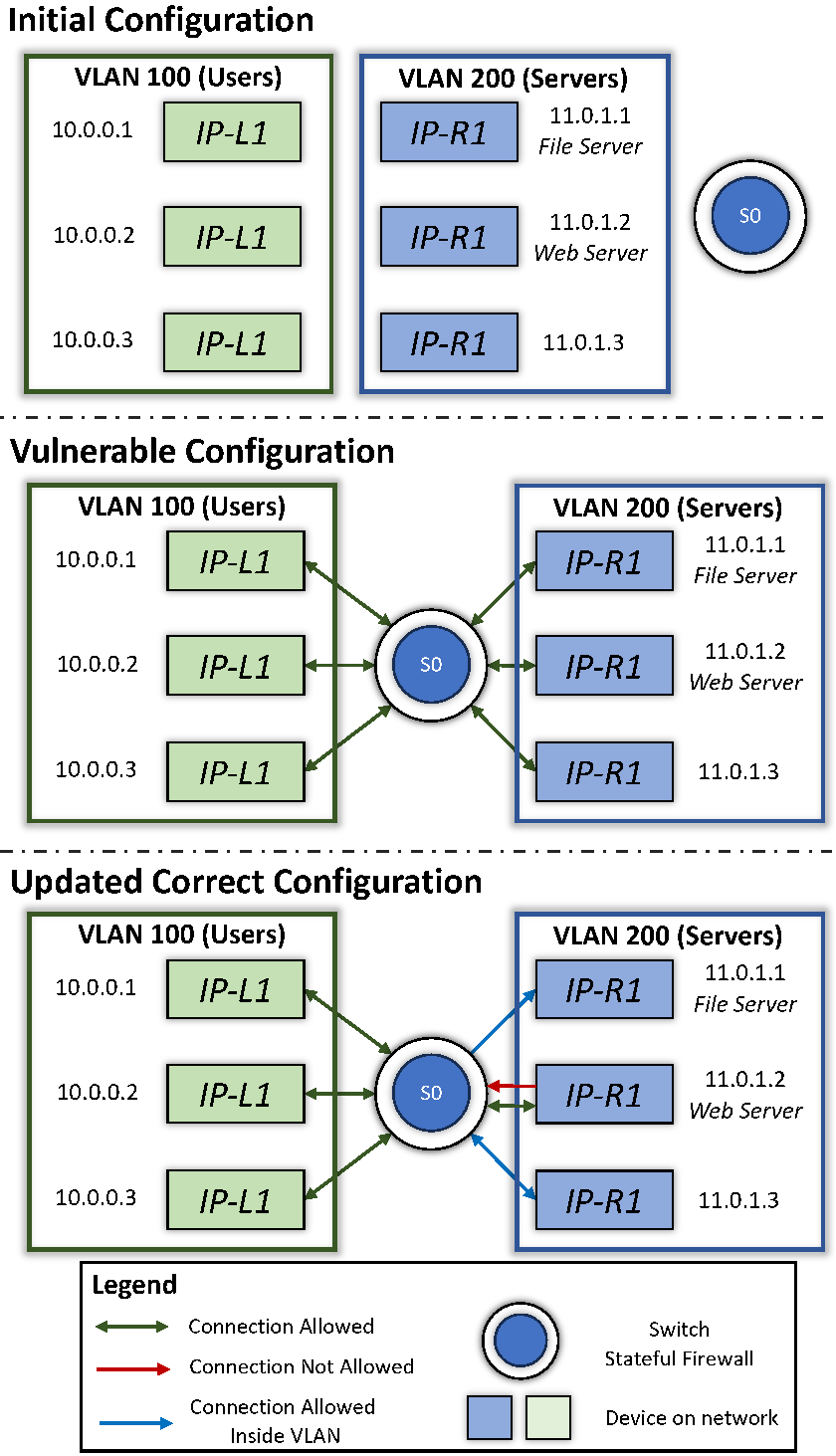}
        \caption{Scenario with missing firewall rules}
        \label{fig:scenario:missing:firewal}
    \end{subfigure}
    \caption{Dataset creation procedure and misconfiguraiton scenarios}
    \label{fig:combined}
\end{figure*}

In our context, a dataset is a file with the same data included in a Network State Change Report (NSCR); in practice, a dataset simulates the result of executing \APPR Steps 1 to 3 and enables assessing \APPR anomaly detection performance. We created three groups of datasets (called real, synthetic, and realistic) following the process depicted in Fig~\ref{fig:datasets:creation} and described below.

% \begin{figure}[ht]
% 	\includegraphics[width=8.4cm]{images/DataSetsCreationProcedure_Updated_Major_Revision.pdf}
%  \caption{Datasets creation procedure}
%  \label{fig:datasets:creation}
% \end{figure}

To create our datasets, we collected from Shodan, a public database with information about public Internet IPs (e.g., open ports, network provider name), the history of port state changes for IPs belonging to \SES. Shodan periodically scans all the IPs on the Internet and keeps track of the open ports; because of such characteristic, Shodan data can be used to simulate the data produced by the NMAP component in \APPR and enable a large-scale assessment of \APPR. 

%Oldest Date: 08/Apr/21
% Latest Date: 04/Feb/23
We analyzed Shodan data collected in the period between April 2021 and May 2023; it concerns 405 SES hosts (we assume each IP belongs to a distinct host). We determine port state changes by comparing port states collected in subsequent scans. We ended up with a change dataset indicating, for every day scanned by Shodan, state change information for every monitored IP. The max number of hosts with port state changes in a day is 13, the min, max, and average number of ports with a modified state are 0, 13, and 1.61, respectively. 

Since Shodan does not monitor all the networks managed by \SES or competing companies (e.g., private corporate networks) it does not enable assessing \APPR with data capturing the characteristics of all the possible contexts in which it could be applied, such as a larger number of host with port state changes, or a type of port change different than the one observed in the Shodan dataset. Further, Shodan does not distinguish between vulnerable and valid port changes.

To address the limitations above, to build our datasets, we extended the data collected by Shodan with additional data, as described below.

% \begin{figure}[t]
% 	\hspace{10mm}\includegraphics[width=7cm]{images/Example 1(redundent rule).pdf}
%  \caption{Scenario with redundant rule}
%        \label{fig:scenario:redundant}
% \end{figure}

% \begin{figure}[t]
% 	\hspace{10mm}\includegraphics[width=7cm]{images/Example 2(missing flow rules).pdf}
%  \caption{Scenario with missing firewall rules}
%  \label{fig:scenario:missing:firewal}
% \end{figure}

\vspace{1em}
\emph{Synthetic datasets.}
We aim at ensuring that four scenarios considered relevant by our industry partner \SES are detected by \APPR. Three scenarios are similar to the one depicted in Fig.~\ref{fig:approach:topology}, where a flow rule is not deleted in a reconfiguration; these three scenarios differ for the port targeted by the flow rule (one for NTP port 123, one for SMB port 445, and one for HTTP port 8080). For clarity, Fig.~\ref{fig:scenario:redundant} shows what would be the correct updated configuration in this case, and what is, instead, the vulnerable updated configuration. The fourth scenario (Fig.~\ref{fig:scenario:missing:firewal}) captures a situation where, in the updated configuration, a subnet is connected to a switch but a stateful firewall that should enable only the WebServer in the subnet to be accessed from the outside is not configured.

We created 100 datasets including configuration changes belonging to the four reference scenarios. 
To this end, we first relied on virtual machines to create the four reference scenarios. For all of them, we used \APPR (Steps 1 and 2) to test both the vulnerable reconfiguration and a valid reconfiguration; then we generated the NSCR (Step 3).

We populated 100 datasets as follows. First, we simulated, in each dataset, 72 hosts with a correct reconfiguration, by copying entries randomly selected from the NSRCs belonging to valid reconfigurations; the number 72 results from duplicating 12 times the 6 valid changes in the monitored scenarios.
Please note that, to simulate different hosts, we replace the IP address of the copied entries with an IP not appearing yet in the dataset\footnote{We repeat such procedure every time we copy an entry, also for building the other datasets; for brevity, we will not report such clarification further. Whenever we indicate that we copy an entry, we mean that we copy the entry and update the IP address appropriately.}.
Then, we introduced into the datset a number of vulnerable hosts by duplicating entries for hosts with vulnerable changes (we took them from the the NSCRs belonging to vulnerable reconfigurations). 
Since the number of vulnerable hosts in a system may vary, we constructed 100 datasets, each having between 5 and 50 vulnerable hosts (we generated 10 datasets for each number of vulnerable hosts in the range, in steps of five).
%sampled from 72 resulting from 
Finally, in each dataset, we reached a total of 405 entries by duplicating entries without state changes from the NSCRs belonging to correct reconfigurations (we updated the IP addresses to match SES hosts).

\vspace{1em}
\emph{Real datasets.} 
To assess \APPR with real data, we generated 20 datasets using the data collected by Shodan; we considered the first available date for each IP. Based on Shodan's data, these datasets include 10 entries with correct configuration changes (the number of modified ports range from 5 to 50) and 395 entries without any change. Then, to obtain datasets with diverse vulnerable entries, we introduced, in each of the 20 dataset, a number of entries with vulnerable configurations by replicating the process adopted for the synthetic datasets (i.e., copying the vulnerable entries from the synthetic NSCRs) but considering a number of violations that varies between 5 and 10; we limit the number of vulnerable entries to 10 because it is very unlikely to observe a higher proportion of vulnerable state changes (with 10 correct changes, 10 vulnerable changes account for 50\%). 

\vspace{1em}
\emph{Realistic datasets.} 
We created 100 datasets to assess \APPR using NSCRs with a number of valid state changes that is larger than what reported in daily Shodan scans, to simulate what happens in private corporate networks not monitored by Shodan.

We constructed each dataset by copying 100 randomly selected entries with changes from the Shodan change dataset. Then, we copy $N$ entries with vulnerable changes from the NSCRs belonging to synthetic scenarios with vulnerable changes. 
We considered a number of $N$ vulnerable entries between 5 and 50, and created 10 dataset files for each, thus ending up with 100 datasets, in total. 

Finally, for each dataset, we sampled and copied additional entries with no state change from the Shodan dataset, till we reached a total of 405 entries in each file. 
%The number of valid changes in our realistic dataset thus ranged from XX to YY (first, second, and third quartile are ZZ, GG, HH).

%To assess \APPR, we conducted our experiments involving the analysis of the generated datasets on the HPC of the University of Luxembourg, with two cores ..

%\subsection{Experiment Setup}

% explain the  construction of network topologies using VM's and mininet along with opendaylight

% explain the four network scenarios i..e, NTP, SMB, HTTP, and HTTP running behind firewall, should all the picture of topologies when explaining

% specs of the machine?

%For RQ3 ...

\subsection{RQ1: Pruning effectiveness}

\subsubsection{Experiment Design}

RQ1 concerns the effectiveness of \APPR's pruning solution (i.e., removal of entries with no state changes). To address RQ1, we compared the results obtained by \APPR pipelines with pruning enabled and \APPR pipelines without pruning. Precisely, since our datasets simulate the results of \APPR Steps 1 to 3, the evaluated pipelines consist of \APPR Steps 4 and 5.

%We applied \APPR Step 4 to all the datasets described in Section~\ref{sec:empirical:subjects}. 
We considered all the four different anomaly detection approaches integrated into \APPR: SIF, SHAC, SKM, SKNN; for SKNN, we considered different configurations for K (i.e., 10, 15, 20, and 25). In total, we selected seven \APPR algorithm configurations. 
We executed the selected \APPR anomaly detection algorithm configurations, with and without pruning, ending up with 14 \APPR anomaly detection algorithm executions.
Finally, to assess \emph{\APPR hosts selection}, as stopping condition (SC), we considered a number of false positives ranging from 1 to 10, in steps of 1, and the case in which 20 false positives should be observed before stopping the inspection; in total, we have 11 configurations for SC.
It leads to 154 ($14\times 11$) \APPR pipelines being assessed.
 
Further, as baseline, we considered also cases in which the datasets are processed by the state-of-the art algorithms IF and LOF; for LOF we considered the same values for K considered for SKNN (i.e., 10, 15, 20, and 25). This leads to 5 additional pipelines, for a total of 159 anomaly detection pipelines assessed in our experiments.

To deal with randomness, we executed each anomaly detection pipeline 40 times on each dataset. Since, in total, we have 220 datasets, it leads to a total of 1.399.200 runs ($159\times40\times220$).

\APPR produces as output a prioritized NSCR, with hosts being prioritized according to their anomaly score (highest first). Since, for each dataset, we know what are the vulnerable hosts, we can determine false positives (i.e., hosts inspected before reaching the stopping condition that are not vulnerable) and true positives (i.e., hosts inspected before reaching the stopping condition that are vulnerable). 

To assess \emph{\APPR hosts selection}, given a \APPR's prioritized NSCR, we can determine the number of false and true positive observed. 
%For \APPR pipelines, we  basically count the number of false and true positives observed till the stopping condition is met. 
False and true positives enable us to compute precision, recall, and F1 score according to standard formula.

To assess \emph{\APPR hosts prioritization}, we count the number of hosts to inspect before identifying all the true positives, which enables determining how quickly a configuration helps engineers detect all the faults. We refer to such metric as \emph{debugging cost (DC)}. By definition, \APPR pipelines differing only for SC, will lead to the same debugging cost.

% Preprocessing improves \APPR results if it leads to better performance measures (i.e., lower debugging cost, and higher precision, recall, and F1 score).  For each considered performance metric, we compare the best performance result obtained with and without preprocessing.  

%Since preprocessing may be effective for certain \APPR algorithm configurations but not others, for each \APPR algorithm configuration, we compare the best performance result obtained with and without preprocessing. 
To perform our assessment, we consider each dataset separately because different \APPR pipelines may perform differently with different datasets. 
%Further, since the real dataset includes only 5 or 10 vulnerable hosts out of 405 hosts, we discuss also the results obtained with the realistic and synthetic datasets, for NSCRs with 5 or 10 vulnerable hosts; we refer to them as \emph{realistic-5} and \emph{synthetic-5}.

To determine if, overall, pruning is beneficial for \APPR, we compare the best result obtained with any \APPR pipeline, in each dataset, with and without pruning. Pruning is beneficial if it enables achieving better metrics in all the datasets.

\subsubsection{Results}

% \usepackage[normalem]{ulem}
\begin{table*}[]
\caption{Results for synthetic, real, and realistic dataset with and without pruning}
\label{Table:result:combined}
% {\fontsize{3.5pt}{2pt}\selectfont
\tiny
\centering
\begin{tabular}{
@{\hspace{1pt}}>{\raggedright\arraybackslash}p{9mm}|
@{\hspace{1pt}}>{\raggedleft\arraybackslash}p{5mm}
@{\hspace{1pt}}>{\raggedleft\arraybackslash}p{5mm}
@{\hspace{1pt}}>{\raggedleft\arraybackslash}p{5mm}
@{\hspace{1pt}}>{\raggedleft\arraybackslash}p{5mm}
@{\hspace{1pt}}>{\raggedleft\arraybackslash}p{5mm}
@{\hspace{1pt}}>{\raggedleft\arraybackslash}p{5mm}
@{\hspace{1pt}}>{\raggedleft\arraybackslash}p{5mm}
@{\hspace{1pt}}>{\raggedleft\arraybackslash}p{5mm}
|
@{\hspace{1pt}}>{\raggedleft\arraybackslash}p{5mm}
@{\hspace{1pt}}>{\raggedleft\arraybackslash}p{5mm}
@{\hspace{1pt}}>{\raggedleft\arraybackslash}p{5mm}
@{\hspace{1pt}}>{\raggedleft\arraybackslash}p{5mm}
@{\hspace{1pt}}>{\raggedleft\arraybackslash}p{5mm}
@{\hspace{1pt}}>{\raggedleft\arraybackslash}p{5mm}
@{\hspace{1pt}}>{\raggedleft\arraybackslash}p{5mm}
@{\hspace{1pt}}>{\raggedleft\arraybackslash}p{5mm}
|
@{\hspace{1pt}}>{\raggedleft\arraybackslash}p{5mm}
@{\hspace{1pt}}>{\raggedleft\arraybackslash}p{5mm}
@{\hspace{1pt}}>{\raggedleft\arraybackslash}p{5mm}
@{\hspace{1pt}}>{\raggedleft\arraybackslash}p{5mm}
@{\hspace{1pt}}>{\raggedleft\arraybackslash}p{5mm}
@{\hspace{1pt}}>{\raggedleft\arraybackslash}p{5mm}
@{\hspace{1pt}}>{\raggedleft\arraybackslash}p{5mm}
@{\hspace{1pt}}>{\raggedleft\arraybackslash}p{5mm}
@{\hspace{1pt}}>{\raggedleft\arraybackslash}p{5mm}
@{\hspace{1pt}}>{\raggedleft\arraybackslash}p{5mm}
}
\hline
&
\multicolumn{8}{c}{\textbf{Synthetic}}                                                                                                                                                                            & \multicolumn{8}{|c}{\textbf{Real}}                                                                                                                & \multicolumn{8}{|c}{\textbf{Realistic}}                                                                                                             \\ \hline
\textbf{}          & \multicolumn{4}{c}{\textbf{Without Pruning}}                                                 & \multicolumn{4}{|c}{\textbf{With Pruning}}                               & \multicolumn{4}{|c}{\textbf{Without Pruning}}                                     & \multicolumn{4}{|c}{\textbf{With Pruning}}                    & \multicolumn{4}{|c}{\textbf{Without Pruning}}                                      & \multicolumn{4}{|c}{\textbf{With Pruning}}                     \\ \hline
\textbf{Algorithm} & \textbf{SC} & \textbf{Prec.}      & \textbf{Rec.}       & \multicolumn{1}{c|}{\textbf{DC}}    & \textbf{SC} & \textbf{Prec.}      & \textbf{Rec.}       & \textbf{DC}    & \textbf{SC} & \textbf{Prec.} & \textbf{Rec.} & \multicolumn{1}{c|}{\textbf{DC}}   & \textbf{SC} & \textbf{Prec.} & \textbf{Rec.} & \textbf{DC}    & \textbf{SC} & \textbf{Prec.} & \textbf{Rec.} & \multicolumn{1}{c|}{\textbf{DC}}    & \textbf{SC} & \textbf{Prec.} & \textbf{Rec.} & \textbf{DC}    \\ \hline \hline
SIF               & 1           & \textbf{0,94}       & 0,88                & \multicolumn{1}{c|}{41,11}          & 1           & \textbf{0,94}       & 0,90                & \textbf{35,48} & 20          & \textbf{0,32}  & 1,00          & \multicolumn{1}{c|}{\textbf{9,99}} & 20          & \textbf{0,42}  & 1,00          & \textbf{14,49} & 20          & \textbf{0,37}  & 0,96          & \multicolumn{1}{c|}{\textbf{46,52}} & 20          & \textbf{0,39}  & 0,93          & 53,91          \\ \hline
SIF              & 20          & 0,50                & \textbf{0,94}       & \multicolumn{1}{c|}{41,11}          & 20          & 0,50                & \textbf{0,94}       & \textbf{35,48} & 20          & 0,32           & \textbf{1,00} & \multicolumn{1}{c|}{\textbf{9,99}} & 20          & 0,42           & \textbf{1,00} & \textbf{14,49} & 20          & 0,37           & \textbf{0,96} & \multicolumn{1}{c|}{\textbf{46,52}} & 20          & 0,39           & \textbf{0,93} & 53,91          \\ \hline
IF                 & N/A         & \textbf{0,00}       & \textbf{0,00}       & \multicolumn{1}{c|}{N/A}            & N/A         & \textbf{0,06}       & 0,17                & N/A            & N/A         & \textbf{0,00}  & 0,34          & \multicolumn{1}{c|}{N/A}           & N/A         & \textbf{0,59}  & 0,97          & N/A            & N/A         & \textbf{0,00}  & 0,00          & \multicolumn{1}{c|}{N/A}            & N/A         & \textbf{0,53}  & 0,51          & N/A            \\ \hline
SHAC               & 1           & \textbf{0,92}       & 0,75                & \multicolumn{1}{c|}{247,88}         & 1           & \textbf{0,92}       & 0,77                & 50,00          & 1           & \textbf{0,38}  & 0,50          & \multicolumn{1}{c|}{10,40}         & 20          & \textbf{0,42}  & 1,00          & 15,00          & 20          & \textbf{0,00}  & 0,00          & \multicolumn{1}{c|}{353,53}         & 1           & \textbf{0,94}  & 0,84          & 47,80          \\ \hline
SHAC               & 20          & 0,46                & \textbf{0,76}       & \multicolumn{1}{c|}{247,88}         & 20          & 0,46                & \textbf{0,80}       & 50,00          & 20          & 0,32           & \textbf{1,00} & \multicolumn{1}{c|}{10,40}         & 20          & 0,42           & \textbf{1,00} & 15,00          & 20          & 0,00           & \textbf{0,00} & \multicolumn{1}{c|}{353,53}         & 20          & 0,48           & \textbf{0,87} & 47,80          \\ \hline
SKM                & 1           & \textbf{0,92}       & 0,76                & \multicolumn{1}{c|}{298,66}         & 1           & \textbf{0,92}       & 0,69                & 60,79          & 1           & \textbf{0,34}  & 0,45          & \multicolumn{1}{c|}{10,01}         & 20          & \textbf{0,42}  & 1,00          & 15,12          & 20          & \textbf{0,00}  & 0,00          & \multicolumn{1}{c|}{365,52}         & 1           & \textbf{0,94}  & 0,83          & 49,81          \\ \hline
SKM                & 20          & 0,47                & \textbf{0,78}       & \multicolumn{1}{c|}{298,66}         & 20          & 0,43                & \textbf{0,75}       & 60,79          & 20          & 0,32           & \textbf{1,00} & \multicolumn{1}{c|}{10,01}         & 20          & 0,42           & \textbf{1,00} & 15,12          & 20          & 0,00           & \textbf{0,00} & \multicolumn{1}{c|}{365,52}         & 7           & 0,70           & \textbf{0,84} & 49,81          \\ \hline
LOF-10             & N/A         & {\ul \textbf{1,00}} & \textbf{0,83}       & \multicolumn{1}{c|}{N/A}            & N/A         & \textbf{1,00}       & 0,83                & N/A            & N/A         & \textbf{0,32}  & \textbf{0,67} & \multicolumn{1}{c|}{N/A}           & N/A         & \textbf{0,00}  & \textbf{0,00} & N/A            & N/A         & \textbf{0,11}  & \textbf{0,10} & \multicolumn{1}{c|}{N/A}            & N/A         & \textbf{1,00}  & \textbf{0.75} & N/A            \\ \hline
LOF-15             & N/A         & \textbf{0,75}       & \textbf{0.88}       & \multicolumn{1}{c|}{N/A}            & N/A         & \textbf{0,79}       & 0,88                & N/A            & N/A         & \textbf{0,32}  & \textbf{0,67} & \multicolumn{1}{c|}{N/A}           & N/A         & \textbf{0,00}  & \textbf{0,00} & N/A            & N/A         & \textbf{0,07}  & \textbf{0,15} & \multicolumn{1}{c|}{N/A}            & N/A         & \textbf{1,00}  & \textbf{0.70} & N/A            \\ \hline
LOF-20             & N/A         & \textbf{0,62}       & \textbf{0,91}       & \multicolumn{1}{c|}{N/A}            & N/A         & 0,65                & \textbf{0,91}       & N/A            & N/A         & \textbf{0,32}  & \textbf{0,67} & \multicolumn{1}{c|}{N/A}           & N/A         & \textbf{0,00}  & \textbf{0,00} & N/A            & N/A         & \textbf{0,07}  & \textbf{0,20} & \multicolumn{1}{c|}{N/A}            & N/A         & \textbf{1,00}  & \textbf{0.78} & N/A            \\ \hline
LOF-25             & N/A         & \textbf{0,55}       & \textbf{0,93}       & \multicolumn{1}{c|}{N/A}            & N/A         & 0,58                & \textbf{0,93}       & N/A            & N/A         & \textbf{0,32}  & \textbf{0,67} & \multicolumn{1}{c|}{N/A}           & N/A         & \textbf{0,00}  & \textbf{0,00} & N/A            & N/A         & \textbf{0,07}  & \textbf{0,25} & \multicolumn{1}{c|}{N/A}            & N/A         & \textbf{1,00}  & \textbf{0.72} & N/A            \\ \hline
SKNN-10            & 1           & \textbf{0,94}       & 0,87                & \multicolumn{1}{c|}{156,25}         & 1           & \textbf{0,94}       & 0,88                & 59,18          & 20          & \textbf{0,32}  & 1,00          & \multicolumn{1}{c|}{11,25}         & 20          & \textbf{0,42}  & 1,00          & 17,50          & 1           & \textbf{0,00}  & 0,00          & \multicolumn{1}{c|}{363,27}         & 1           & \textbf{0,89}  & 0,65          & 36,31          \\ \hline
SKNN-10            & 20          & 0,47                & \textbf{0,88}       & \multicolumn{1}{c|}{156,25}         & 20          & 0,43                & {\ul \textbf{0,96}} & 59,18          & 20          & 0,32           & \textbf{1,00} & \multicolumn{1}{c|}{11,25}         & 20          & 0,42           & \textbf{1,00} & 17,50          & 1           & 0,00           & \textbf{0,00} & \multicolumn{1}{c|}{363,27}         & 20          & 0,48           & \textbf{0,98} & 36,31          \\ \hline
SKNN-15            & 1           & \textbf{0,94}       & 0,92                & \multicolumn{1}{c|}{56,01}          & 1           & {\ul \textbf{0,95}} & 0,93                & 36,44          & 20          & \textbf{0,32}  & 1,00          & \multicolumn{1}{c|}{10,85}         & 20          & \textbf{0,42}  & 1,00          & 15,50          & 1           & \textbf{0,00}  & 0,00          & \multicolumn{1}{c|}{363,26}         & 1           & \textbf{0,90}  & 0,66          & 30,92          \\ \hline
SKNN-15            & 20          & 0,50                & {\ul \textbf{0,96}} & \multicolumn{1}{c|}{56,01}          & 20          & 0,51                & \textbf{0,95}       & 36,44          & 20          & 0,32           & \textbf{1,00} & \multicolumn{1}{c|}{10,85}         & 20          & 0,42           & \textbf{1,00} & 15,50          & 1           & 0,00           & \textbf{0,00} & \multicolumn{1}{c|}{363,26}         & 20          & 0,51           & \textbf{1,00} & 30,92          \\ \hline
SKNN-20            & 1           & \textbf{0,94}       & 0,92                & \multicolumn{1}{c|}{37,85}          & 1           & \textbf{0,94}       & 0,89                & 37,98          & 20          & \textbf{0,32}  & 1,00          & \multicolumn{1}{c|}{10,85}         & 20          & \textbf{0,42}  & 1,00          & 15,50          & 1           & \textbf{0,00}  & 0,00          & \multicolumn{1}{c|}{361,53}         & 1           & \textbf{0,91}  & 0,66          & \textbf{30,65} \\ \hline
SKNN-20            & 20          & 0,50                & \textbf{0,96}       & \multicolumn{1}{c|}{37,85}          & 20          & 0,50                & \textbf{0,92}       & 37,98          & 20          & 0,32           & \textbf{1,00} & \multicolumn{1}{c|}{10,85}         & 20          & 0,42           & \textbf{1,00} & 15,50          & 1           & 0,00           & \textbf{0,00} & \multicolumn{1}{c|}{361,53}         & 20          & 0,51           & \textbf{1,00} & \textbf{30,65} \\ \hline
SKNN-25            & 1           & \textbf{0,94}       & 0,92                & \multicolumn{1}{c|}{\textbf{34,91}} & 1           & \textbf{0,94}       & 0,86                & 41,19          & 20          & \textbf{0,32}  & 1,00          & \multicolumn{1}{c|}{10,85}         & 20          & \textbf{0,42}  & \textbf{1,00} & 15,50          & 1           & \textbf{0,00}  & 0,00          & \multicolumn{1}{c|}{361,63}         & 1           & \textbf{0,91}  & 0,66          & 31,19          \\ \hline
SKNN-25            & 20          & 0,50                & {\ul \textbf{0,96}} & \multicolumn{1}{c|}{\textbf{34,91}} & 20          & 0,49                & \textbf{0,88}       & 41,19          & 20          & 0,32           & \textbf{1,00} & \multicolumn{1}{c|}{10,85}         & 20          & 0,42           & \textbf{1,00} & 15,50          & 1           & 0,00           & \textbf{0,00} & \multicolumn{1}{c|}{361,63}         & 20          & 0,50           & \textbf{1,00} & 31,19          \\ \hline  
\end{tabular}

% }
\end{table*}
% \input{tables/Result_Synthetic_all_Anomaly_groups.tex}

% \input{tables/Result_Real_5n10_Anomaly_groups.tex}

% \input{tables/Result_Realistic_all_Anomaly_groups.tex}

% Table~\ref{} provides descriptive statistics concerning the percentage of entries removed by each dataset when applying pre-processing. The dataset with the largest proportion of entries being removed is ....

Tables~\ref{Table:result:combined}
%~\ref{Table:result:synthetic:five}, and~\ref{Table:result:realistic:five} 
report results obtained by the \APPR pipelines having the highest precision, recall, and F1 score, for each \APPR algorithm. Precisely, for each anomaly detection algorithm, we report on the SC leading to the highest precision, recall, and F1 score (best values are bold, algorithms where SC is not applicable have all the values bold). We underline the best performance value obtained in each dataset. 
We also report the debugging cost (DC) of each algorithm (best dataset value in bold). 

For \emph{hosts selection}, the best results observed with the synthetic dataset without pruning are 1.00 (precision), 0.96 (recall), 0.93 (F1 score); with pruning, we observe, 1.00 (precision), 0.96 (recall), 0.94 (F1 score). Pruning slightly improves F1 score (0.93 VS 0.94), the other metrics are the same.
The best results for the real dataset without pruning are 0.38 (precision), 1.00 (recall), 0.48 (F1 score); with pruning, we observe, 0.59 (precision), 1.00 (recall), 0.59 (F1 score). Pruning improves precision (0.38 VS 0.59) and F1 score (0.48 VS 0.59), recall is maximal in both.
The best results for the realistic dataset without pruning are 0.37 (precision), 0.96 (recall), 0.53 (F1 score); with pruning, we observe, 1.00 (precision), 1.00 (recall), 0.88 (F1 score). Pruning improves precision (0.37 VS 1.00), recall (0.96 VS 1.00) and F1 score (0.53 VS 0.88). Concluding, we can affirm that pruning improves \APPR hosts selection results.

% Focusing on precision, recall, and F1-score, we can notice that for most \APPR pipeline configurations, preprocessing leads to better results (e.g., for SIF, with stopping condition 1, precision is higher with preprocessing, as shown in the top left result column of Tables~\cite{} and~\cite{}). The percentage of performance measurements that are higher for preprocessed datasets are 50\%, 55.6\%, 96.3\%, 100\%, 88.9\%, for synthetic, synthetic-5, realistic, realistic-5, and the real dataset, respectively. Preprocessing may improve a lot the performance of algorithms, with KNN's recall for realistic dataset going from 0.00 to 1.00. The least improvements are  shown for the synthetic dataset ...
% If we look at the best results for the three metrics above, we can notice that preprocessing always leads to better results except for recall with the synthetic dataset, were recall is 0,964 without preprocessing and 0,955 with preprocessing. 

For \emph{hosts prioritization}, we focus on DC.
The best result (i.e., lowest average number of hosts to be inspected) for the synthetic dataset without pruning is 34.91, with pruning is 35.48; it shows that without pruning, engineers inspect 0.57 anomalies less, on average. For the real dataset, the best result without pruning is 9.99, with pruning is 14.49; it shows that without pruning, engineers inspect 4.5 anomalies less, on average. For the realistic dataset, the best result without pruning is 46.52, with pruning is 30.65; it shows a trend that differs from the other two datasets, indeed, without pruning, engineers inspect 15.96 anomalies more, on average. Although in two datasets the lack of pruning slightly reduced the number of anomalies to be inspected, overall, because of what observed in the realistic dataset, pruning leads to a higher reduction of debugging cost. Indeed, considering all the datasets, \APPR without pruning leads to 25.23 DC, while \APPR with pruning leads to 18.47 DC.
%
%(34,91*20*40+9,99*20*40+46,52*100*40)/(220*40)=25,23
%
%(35.48*20*40+14.49*20*40+30.65*100*40)/(220*40)=18,47
%

Concluding, we suggest relying on pruning to perform anomaly detection with \APPR.

% we can observe that preprocessing reduces them for all the \APPR algorithms except for KNN-20 and KNN-25 with the synthetic dataset, SIF with the realistic dataset, and all the lagorithms for the real dataset. However, except for the real dataset, the debugging cost for the \APPR algorithm leading to the best result is always lower with preprocessing except for one dataset.

% Concluding, preprocessing tends to lead to better performance results, but sometimes it may also lead to slightly worser results. However, if KNN is adopted, preprocessing is strongly suggested because without preprocessing we observed that the algorithm may not enable the detection of any anomaly for test selection or have a high debugging cost for test prioritization.

% Since performance deteriorates only for recall in one dataset, 

% We discuss the reported data in the following sections.

\subsection{RQ2: Best \APPR pipeline}

\subsubsection{Experiment Design}

We rely on the same data collected for RQ1. But, since RQ1 shows that pruning leads to best results, we focus on anomaly detection pipelines with pruning.
%We compare the best performance results obtained by each pipeline, in each dataset, to identify the pipeline leading to the best results in most of the datasets.

For hosts selection, the best pipelines for a dataset are the ones that maximize recall (and significantly differ from other pipelines) and, among the ones with highest recall, maximize precision (and significantly differ from other pipelines). Alternatively, F1 score can be considered to identify the best pipelines.
\MAJOR{5.2}{Since hosts selection leads to the inspection of a subset of hosts, potentially overlooking vulnerabilities,} \MAJOR{3.4}{maximizing recall (i.e., the proportion of faulty configurations detected) is a necessity because engineers aim at detecting all the vulnerabilities, to have a secure system. Since precision is the proportion of inspected hosts that are faulty, maximizing precision implies that engineers minimize the inspection of hosts that are valid. Consequently,
once recall is maximized, the approach that maximizes precision is the one that enable using engineers time most effectively.}

For hosts prioritization, the best pipelines for a dataset are the ones minimizing DC (i.e., they help identifying all the vulnerabilities with the minimal number of anomalies to be inspected) and showing significant difference from the others.

Ideally, the best pipelines (i.e., the ones that we suggest for their use with \APPR) are the ones belonging to the intersection of the sets of best pipelines observed with the different datasets.

We compute the significance of the difference between results observed with datasets having the same number of anomalies by computing p-values with the Mann–Whitney U test~\cite{Utest}, a non parametric method; we report a significant difference when p-values are below 0.05. 
%However, since our metrics might be affected by the number of anomalies in the dataset (e.g., it's easier to achieve a recall equal to 1.00 with only few vulnerable hosts), the significance of the different is computed for results obtained from datasets having the same number of vulnerabilities. Consequently, for the realistic and synthetic dataset, for each pair of pipelines, we compute ten different p-values (for 5 to 50 anomalies) and the difference is significant if they are all below 0.05.

\subsubsection{Results}

%We identify the best approach by discussing the results obtained in the different datasets. Basically, 
We aim to identify the subset of pipelines that perform best in all the datasets.

For hosts selection, with the synthetic dataset, the best performing approach is SKNN-10 with SC=20 (recall is 0.96); however, it does not significantly differ from SKNN-15 with SC=20 (recall is 0.95) and SKNN-20 with SC=20 (recall is 0.92). Further, SKNN-15 with SC=20 and SKNN-20 with SC=20 have a significantly higher precision. Other approaches whose recall does not significantly differ from 
SKNN-10 with SC=20 are SIF SC=20 (recall is 0.94), LOF 20 (recall is 0.91), and LOF 25 (recall is 0.93). For the real dataset, all the pipelines except the ones based on LOF and IF reach maximum recall. The poorer performance observed for LOF and IF (recall is zero for LOF and below 1 for IF) is likely due to the fact that, since the real dataset includes only few changes (vulnerable and correct), what happens is that all the changes (correct and vulnerable) may look similar (i.e., are very close if we apply a distance metric). Consequently, approaches that simply select a subset of ports based on a threshold on the anomaly score (e.g., LOF) perform much worse than approaches that sort all the anomalies (i.e., SIF, SHAC, SKM, SKNN) and provide some tolerance for classification mistakes (i.e., they use SC=20). Indeed, SC equal to 20 enables \APPR pipelines to include the vulnerable hosts in their selection, instead, lower SC values doesn't (please note that other values for SC do not appear in the table because SC $\le 10$ does not enable selecting any vulnerable host).
%Instead, approaches that select only a small subset of hosts having a large distance from others  (i.e., all the  LOF) cannot identify the vulnerable hosts.
The IF result is nevertheless interesting because, despite not being able to achieve max recall, it has a better precision than other pipelines. However, the pipelines performing best in both the synthetic and the real dataset are SKNN-10 with SC=20, SKNN-15 with SC=20, SKNN-20 with SC=20, and SIF with SC=20. Finally, in the realistic dataset, among SIF and SKNN-based pipelines, the best recall results (0.86 and 0.87) are those obtained by SKNN-based pipelines, with the best precision (0.51) achieved by SKNN-15 with SC=20 and SKNN-20 with SC=20. 
%Also, they perform significantly better than SIF SC=20. 
Concluding the best \APPR pipelines are SKNN-15 with SC=20 and SKNN-20 with SC=20, with SKNN-15 with SC=20 to be favoured since it had slightly better results in the synthetic dataset.

For hosts prioritization, recall that we ignore the configuration value assigned to SC because all hosts are inspected and we aim at determining how quickly all the vulnerabilities are found. With the synthetic dataset, the best result is achieved by SIF with 35.48 hosts to be inspected on average. However, such result does not significantly differ from that of SKNN-15 (36.44) and SKNN-20 (37.98). For the real dataset, SIF still performs the best (14.49), with SKNN-15 and SKNN-20 performing similarly (15.50); also, as observed for hosts selection, clustering algorithms perform well (SHAC's DC is 15, SKM's is 15.12), which indicates that they perform well when only a few state changes are observed. Finally, for the realistic dataset, the best results are observed with SKNN-15 (30.92) and SKNN-20 (30.65), they both perform significantly better than SIF (53.91). Concluding, also for hosts prioritization we suggest relying on SKNN-15 and SKNN-20.

Concluding, our results show that the approach proposed in this paper (i.e., sorting KNN results and relying on thresholds over false positives) outperform state-of-the-art approaches (i.e., IF and LOF). Also, relying on SKNN-15 and SKNN-20 enables achieving best results for both hosts selection and prioritization. \MAJOR{3.4, 3.5}{Since our datasets include vulnerabilities affecting reachability (Scenarios 1, 2, and 3) and confidentiality (Scenarios 1, 2, 3, and 4), the high recall achieved by the best pipeline enables us to conclude that \APPR empower engineers in effectively assessing both security properties.}

\subsection{RQ3 - Accuracy of host matching module}
\label{sec:rqThree}

\subsubsection{Experiment Design}
We aim to assess the accuracy of the host
matching component and its impact on FISTS results. Since Shodan does not provide information about what hosts have changed their IP or MAC address from one scan to the other, we created additional datasets where we introduced changes in the IP and MAC addresses of hosts and verify if \APPR properly process them. Precisely, we followed the process described in Section~\ref{sec:empirical:subjects} to create 100 real datasets with 5 to 50 vulnerable hosts (10 datasets for each different number of vulnerable hosts). From these 100 datasets, \MAJOR{1.5,5.3}{we conducted two experiments (namely EXP1 and EXP2),} deriving 100 
\emph{original NSCR realistic datasets} in each, as follows. \MAJOR{1.5,5.3}{In EXP1, for each real dataset,} we randomly sampled 40 hosts and changed their IP address for the updated configuration, we also sampled 40 other hosts and changed their MAC address. \MAJOR{1.5,5.3}{In EXP2, we randomly sampled 40 hosts and changed both their IP and MAC address.} 
%Then, to ensure having at least 20 hosts with port state changes and IP/MAC changes, we selected 10 hosts with modified IP and 10 hosts with modified MAC, and copied the port state changes observed in the real dataset. 
For each experiment, we thus obtained \emph{modified NSCR realistic datasets}. Finally, for each \emph{modified NSCR realistic dataset}, we derived execution data files for the initial and updated configuration, to be processed by our \emph{host matching and comparison} component. 

Since we know what the expected NSCR is, we can count true and false positives. A true positive is an NSCR entry generated by our algorithm that matches what in the corresponding modified NSCR realistic dataset. A false positive is an NSCR entry not present in the corresponding dataset. We assess our approach in terms of accuracy.

\subsubsection{Results}
The execution of our algorithm on the 100 data file pairs led to perfect accuracy.
%thus demonstrating the effectiveness of the proposed solution.
Further, the distribution of entries having changes in both port state and IP or MAC address,
%concerning Fig~\ref{fig:results:HMC} shows the distribution of entries having changes in both port state and IP or MAC address, within the modified NSCR realistic dataset. The number of those entries, 
in EXP1, ranges within 0 and 35 for each modified NSCR realistic dataset file (median is 12.5), \MAJOR{1.5, 5.3}{in EXP2, it ranges between 22 and 44  (median is 38.5). Given our 100\% accuracy, it demonstrates \APPR HMC algorithm effectiveness with different numbers of port state changes, even in the presence of simultaneous changes of IP and MAC (i.e., EXP2).}

% \begin{figure}[ht]
% \center
% 	\includegraphics[width=5cm]{images/HMC_barplot_new.eps}
%  \caption{Hosts with changes in port state and IP/MAC address.}
%  \label{fig:results:HMC}
% \end{figure}

\subsection{RQ4 - Scalability of approach}
\label{sec:rqFour}

\subsubsection{Experiment Design}
Three are the steps that mainly affect the scalability of the approach: Step 1 - Traffic Generation, Step 2 - Response Monitoring, Step 5 - Vulnerability Prioritization. Executing Step 3 and Step 4 on execution data collected from 405 IPs takes less than two seconds, therefore, we exclude them from our investigation for RQ4.

Since Steps 1 and 2 are coupled (i.e., implemented with NMAP), we assess them together. 
Given that, for security reasons, we cannot scan the SES network directly, we simulated multiple scans of a network with 405 IPs and report on execution time. Precisely, we executed NMAP to independently scan 1000 ports of three hosts in the updated configuration in Fig~\ref{fig:scenario:redundant}; we rely on a simulation with VMs. We repeated the execution 400 times, to collect enough data points (each data point captures the execution time to scan 3 IPs). To simulate multiple scans of a network with 405 IPs, we randomly sampled 135 data points ($135\times3=405$), and repeated the sampling 100 times. We discuss the distribution of execution time, to demonstrate the scalability of \APPR Steps 1 and 2 
\MAJOR{4.4}{in the context of SATCOM providers. Specifically, although SDNs can be quickly reconfigured every few seconds, in our reference context, they are modified based on customer requirements (e.g., companies buying  satellite communication channels), which happens few times in a week, during dedicated management windows of 15 to 30 minutes.}

For Step 5, what affects the scalability of the approach is the time taken by the anomaly detection algorithms. We compare the different algorithms by reporting the time taken to process the different datasets considered for RQ1 and RQ2.

\subsubsection{Results}

%Fig.~\ref{fig:results:nmap_time_405} shows the distribution of 
The time taken by NMAP to scan 1000 ports in 405 IPs ranges between 1816.7 and 1821.6 with a median of 1819.54 seconds (30.32 minutes), which is acceptable since it fits in a typical maintenance window for networks \MAJOR{5.4}{(our industry partner requires monitoring to be performed in such maintenance windows).} \MAJOR{1.8, 5.4}{Further, the scan can be parallelized thus reducing the total time required; however, although we suggest relying on parallel scanning, what may limit its adoption is the network bandwidth dedicated to maintenance operations, which vary across companies.} Last, the time required to scan different sets of IPs is very similar (mix and max are 1816.73s and 1821.62s, respectively), thus suggesting results might be confirmed by repetitions of the experiment in other contexts.

% Minimum: 1816.73
% Maximum: 1821.62
% 1st Quartile (Q1): 1818.86
% 3rd Quartile (Q3): 1820.25
% Median (Q2): 1819.54

% \begin{figure}[ht]
% \center
% 	\includegraphics[width=5cm]{images/1_IP_and_1000_ports.eps}
%  \caption{Time required to scan 1000 ports of 405 IPs; each datapoint corresponds to a distinct simulation (100 in total)}
%  \label{fig:results:nmap_time_405}
% \end{figure}

\begin{figure*}[ht]
\centering
\caption{Execution time of \APPR algorithms}
\label{}
\begin{subfigure}{0.18\textwidth}
	\includegraphics[width=\textwidth]{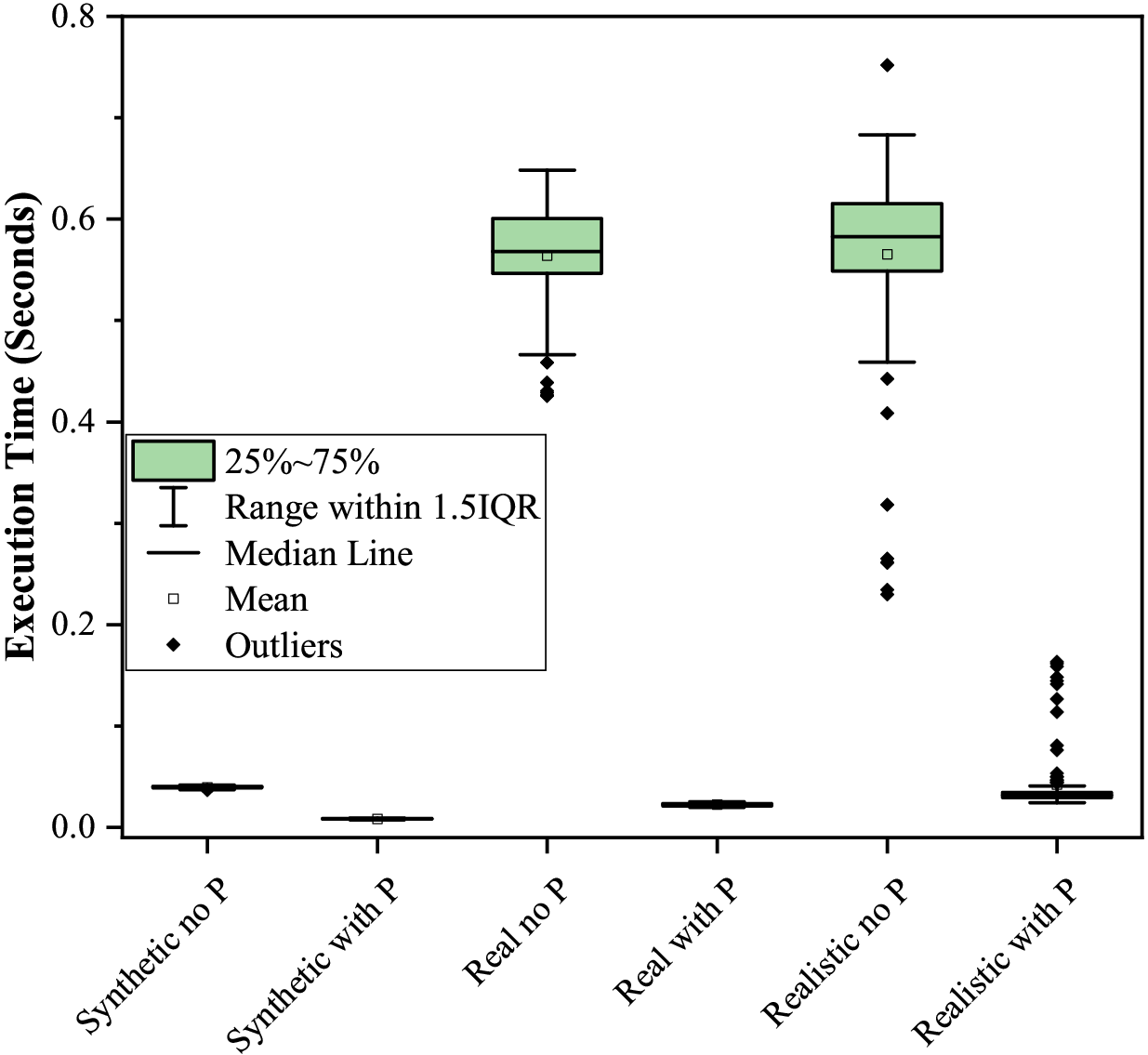}
 \caption{SHAC.}
\label{fig:results:All_Agglomerative}
\end{subfigure}
\begin{subfigure}{0.18\textwidth}
	\includegraphics[width=\textwidth]{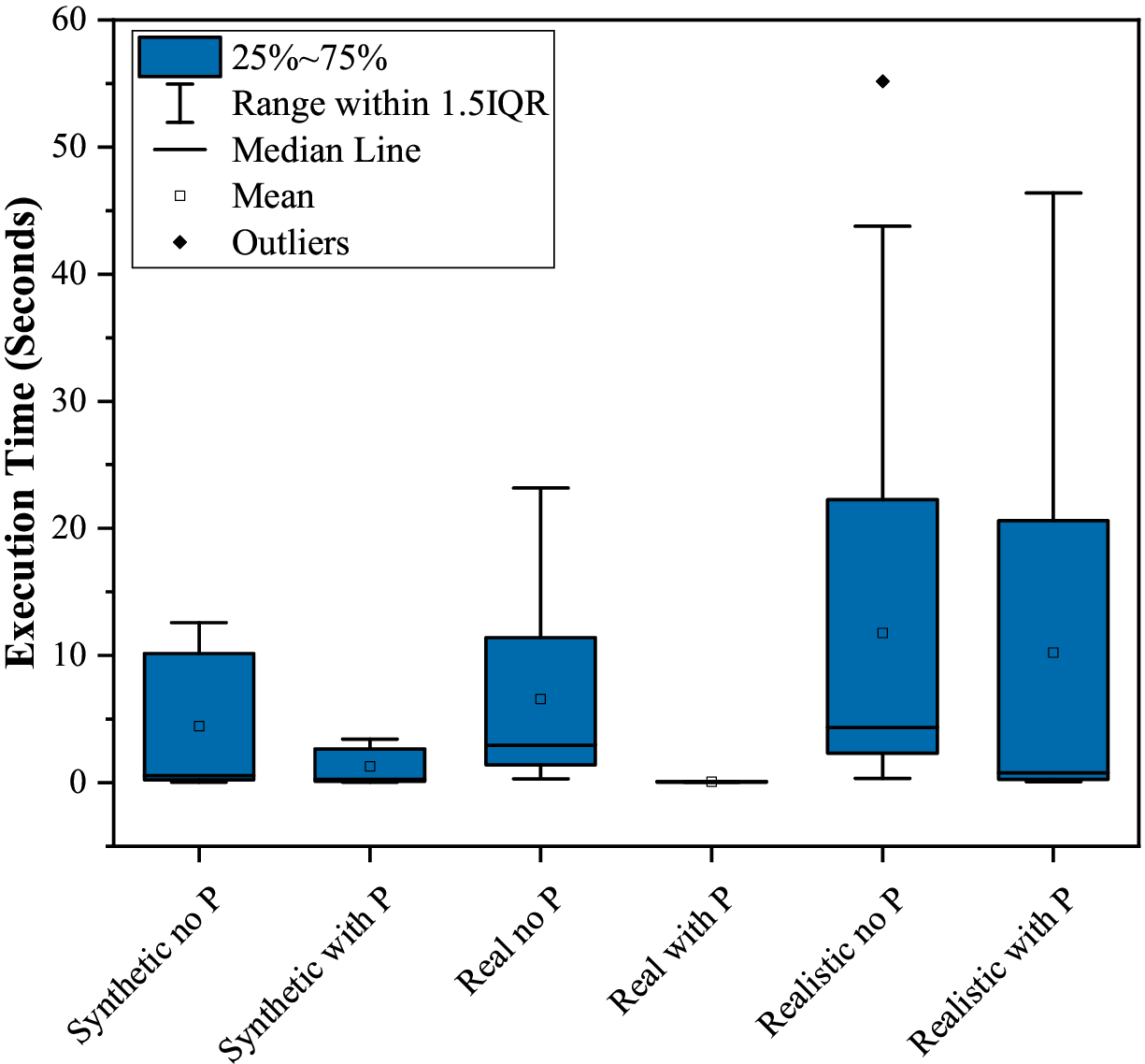}
 \caption{SKM.}
 \label{fig:results:All_k_means} %the label goes BELOW the caption!!!!!
\end{subfigure}
\begin{subfigure}{0.18\textwidth}
 	\includegraphics[width=\textwidth]{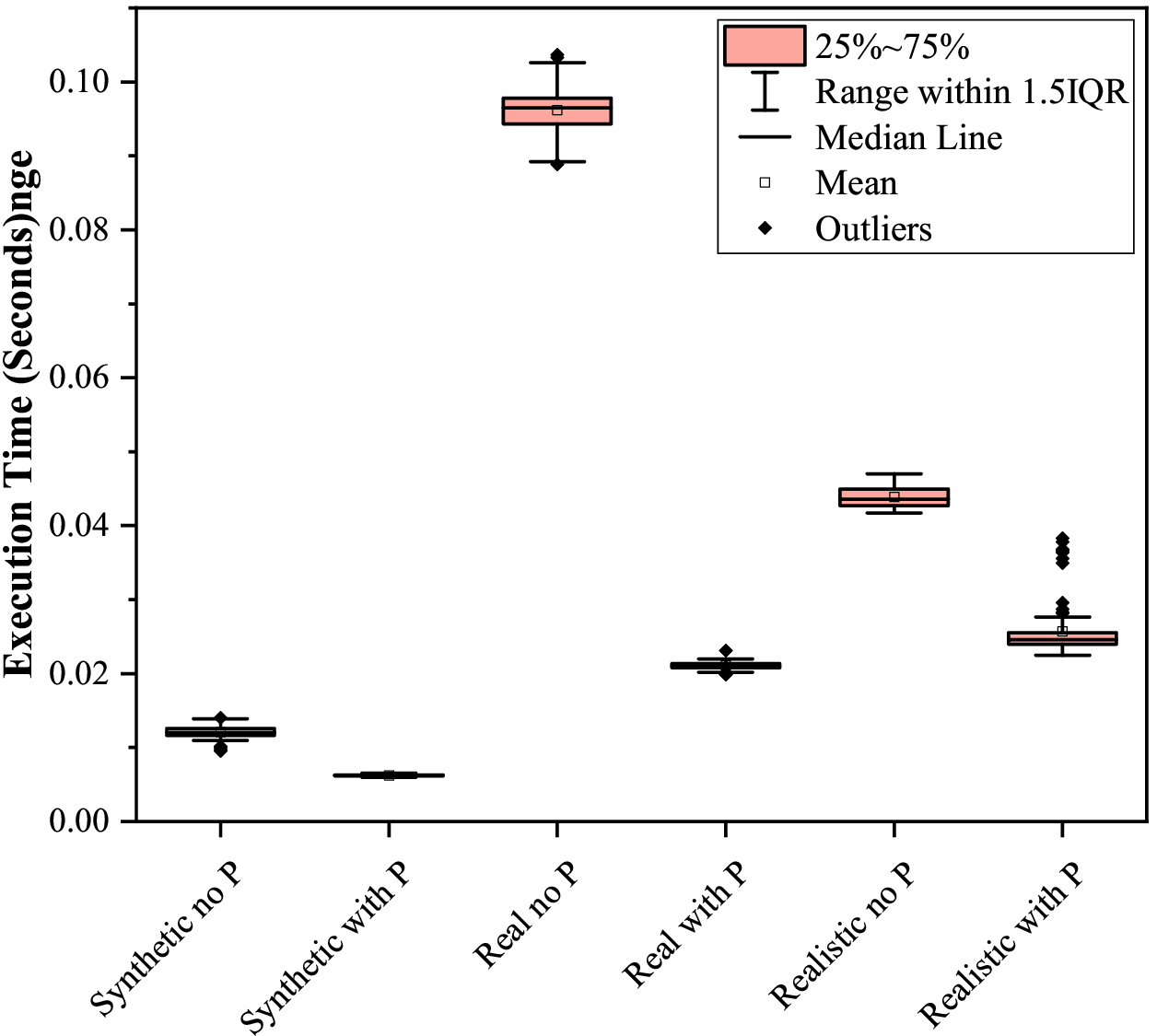}
  \caption{LOF.}
  \label{fig:results:All_LOF}
\end{subfigure}
\begin{subfigure}{0.18\textwidth}
 	\includegraphics[width=\textwidth]{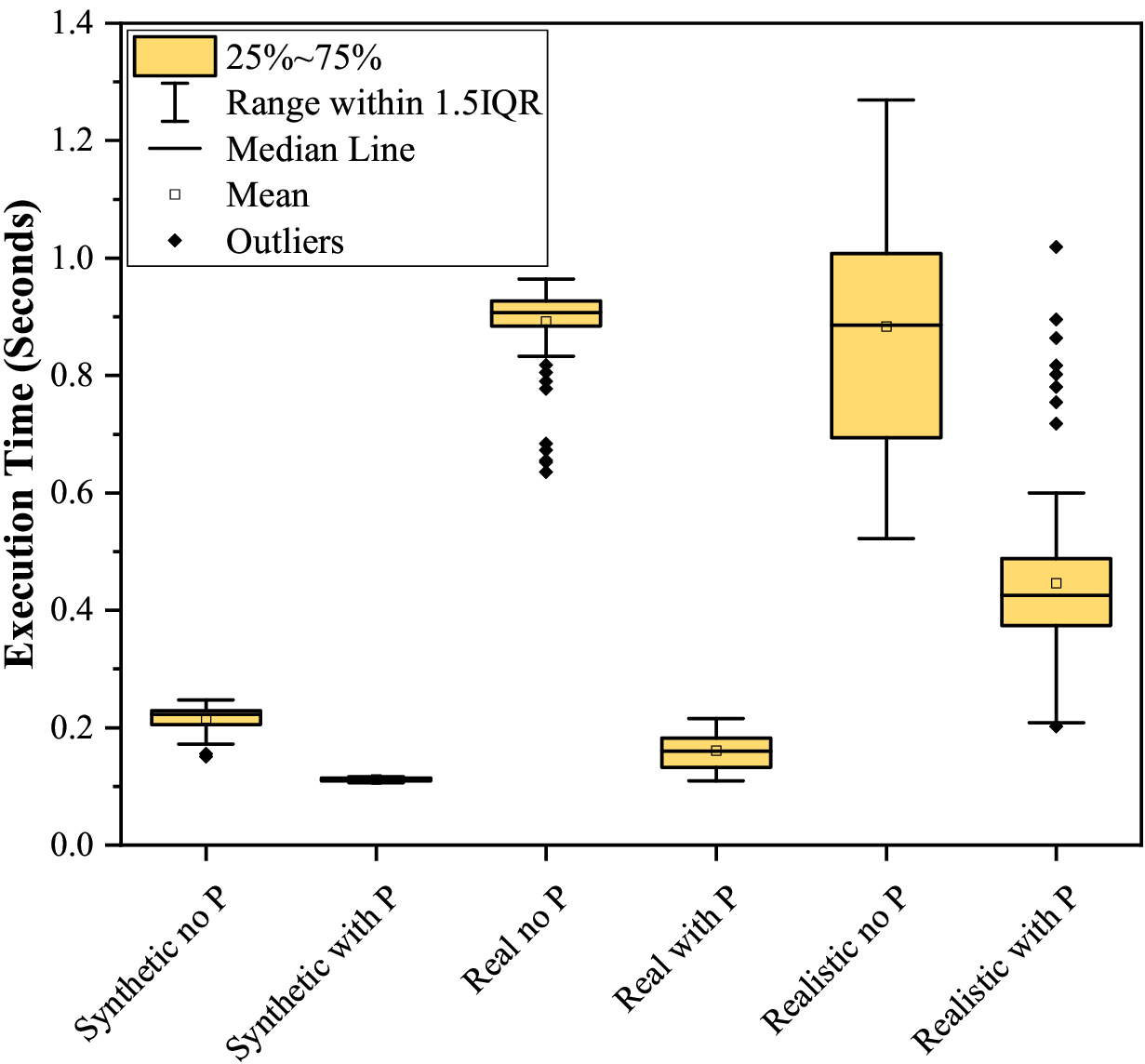}
  \caption{IF/SIF.}
        \label{fig:results:All_IF}
 \end{subfigure}
\begin{subfigure}{0.18\textwidth}
 	\includegraphics[width=\textwidth]{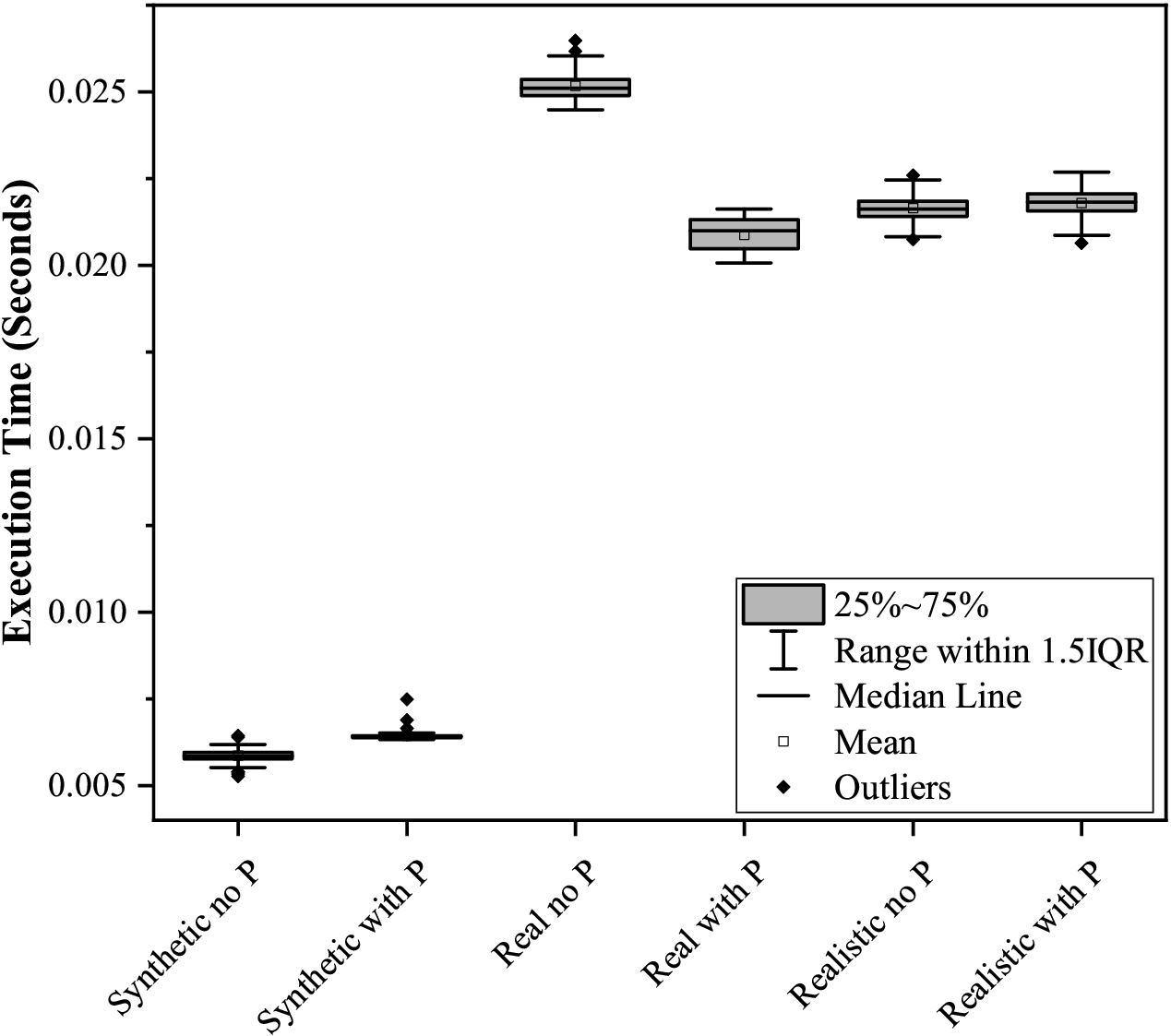}
  \caption{SKNN.}
  \label{fig:results:SKNN}
 \end{subfigure}
\end{figure*}

Figs.~\ref{fig:results:All_Agglomerative} to~\ref{fig:results:SKNN} show the execution time taken by SHAC, SKM, LOF, IF/SIF, and SKNN on all the datasets. SIF and IF take the same execution time because SIF relies on IF, but simply selects all the sorted datapoints instead of a subset.

Our plots show that the use of pruning reduces execution time; indeed, the median execution time observed with pruning (see \emph{with P} in the figures) is lower than what observed without pruning (see \emph{no P} in the figures) in all the cases except the synthetic dataset for SKNN. In the synthetic dataset for SKNN, the execution time is so low (between 0.005 and 0.010 seconds), that the better result obtained without pruning may be due to a slight change in the background tasks running on the machine used for experiments. Concluding, the results observed for execution time provide an additional argument in favor of the adoption of pruning, as we already suggested when addressing RQ1.

All the approaches, except SKM, process each dataset in less than 1.4 seconds, thus showing that they scale well. SKM, instead, takes up to 47 seconds in the case of the realistic dataset with pruning. However, one minute to process one NSCR dataset is an acceptable time since it's an order of magnitude less than what required to collect the data by scanning the network. SKNN, which is the approach selected in RQ2 takes less than 0.3 seconds, thus showing that it is also among the quickest ones.

\subsection{Threats to validity}

For \emph{internal validity}, we  manually inspected the results (i.e., labeled datapoints) for a subset of the datasets. For \emph{construct validity}, we relied on metrics (precision, recall) commonly adopted to assess machine learning algorithms, and a metric (debugging cost) that is an indirect measure of effort. For \emph{conclusion validity}, we relied on non-parametric statistical tests.
\MAJOR{1.11, 5.1}{For generalizability, we constructed datasets that have characteristics (number of hosts, port numbers) similar to those of large SATCOM providers, as confirmed by our industry partners. However, the results (precision, recall) obtained with unsupervised machine learning algorithms may vary across network environments. To address this threat, we considered datasets presenting a varying proportion of hosts with port or IP changes. For vulnerabilities, we constructed scenarios that shall be considered invalid in any context (i.e., messages forwarded to the wrong subnet) but we also considered a scenario that is likely vulnerable (i.e., a whole subnet with only one server becomes fully reachable), based on feedback from SATCOM experts, although it might be considered valid in some peculiar contexts. Concluding, although we expect replicability in SATCOM networks similar to ours, repeatability  in different contexts can't be granted.}
%in highly dynamic and diverse network environments where the definition of "normal" can vary significantly

%% !TEX root =  MAIN.tex
%\clearpage

% \section{Data Availability}

% Our replication package and results are available online~\cite{replicability}.

\section{Conclusion}
\label{sec:conclusion}

In this paper, we addressed the problem of automatically detecting configuration updates for software defined networks (SDNs) that introduce vulnerabilities. Model-based approaches that process configuration files of SDN software are unlikely applicable in industry because commercial software with proprietary formats is used. Therefore, we propose FIeld-based Security Testing
of SDN Configurations Updates (\APPR), a black-box, field-based testing approach that collects data about the port states of hosts on the network before and after a configuration update, and then relies on an anomaly detection algorithm to sort hosts based on the probability of being vulnerable. Also, we propose a strategy to stop inspecting hosts when there is evidence that not inspected hosts are unlikely vulnerable. Further, we integrated a solution to determine if a host present in the initial configuration changed IP or MAC address as a result of the configuration update. Finally, we also suggest an anomaly detection approach (SKNN) that relies on sorting the results produced by the K-Nearest Neighbour algorithm to identify anomalous hosts. 

Our empirical assessment of \APPR shows that best results are obtained with SKNN and pruning, with precision, recall, and F1 score reaching peaks of 0.95, 1.00, and 0.94, respectively. Also, they enable detecting all the anomalies injected in our datasets (up to 50) by inspecting up to 37.98 hosts on average. Further, we demonstrated the accuracy of the host matching algorithm, and the scalability of \APPR, which can scan 405 IPs in 30 minutes (parallelizable), and processes the collected data in less than a minute. Our replication package and results are available online~\cite{replicability}.

%\MAJOR{1.11, 5.4}{Our work focused on the assessment of FISTS with well-known anomaly detection algorithms; future work will include extensive evaluation with additional anomaly detection algorithms along with an assessment of the improvements of benefits obtained relying on semi-supervised learning with historical data.}

\section*{Acknowledgments} 
This work has been supported by SES~\cite{SES} and  the Luxembourg National Research Fund (FNR) under the project INSTRUCT (IPBG19/14016225/INSTRUCT~\cite{IPBG}).

%\clearpage

\bibliographystyle{IEEEtran}
\bibliography{Ref}

\begin{thebibliography}{10}
\providecommand{\url}[1]{#1}
\csname url@samestyle\endcsname
\providecommand{\newblock}{\relax}
\providecommand{\bibinfo}[2]{#2}
\providecommand{\BIBentrySTDinterwordspacing}{\spaceskip=0pt\relax}
\providecommand{\BIBentryALTinterwordstretchfactor}{4}
\providecommand{\BIBentryALTinterwordspacing}{\spaceskip=\fontdimen2\font plus
\BIBentryALTinterwordstretchfactor\fontdimen3\font minus \fontdimen4\font\relax}
\providecommand{\BIBforeignlanguage}[2]{{%
\expandafter\ifx\csname l@#1\endcsname\relax
\typeout{** WARNING: IEEEtran.bst: No hyphenation pattern has been}%
\typeout{** loaded for the language `#1'. Using the pattern for}%
\typeout{** the default language instead.}%
\else
\language=\csname l@#1\endcsname
\fi
#2}}
\providecommand{\BIBdecl}{\relax}
\BIBdecl

\bibitem{SATCOM}
O.~Kodheli, E.~Lagunas, N.~Maturo, S.~K. Sharma, B.~Shankar, J.~F.~M. Montoya, J.~C.~M. Duncan, D.~Spano, S.~Chatzinotas, S.~Kisseleff, J.~Querol, L.~Lei, T.~X. Vu, and G.~Goussetis, ``Satellite communications in the new space era: A survey and future challenges,'' \emph{IEEE Communications Surveys \& Tutorials}, vol.~23, no.~1, pp. 70--109, 2021.

\bibitem{aviation}
{SES Luxembourg}, ``{SAT-based} business-aviation internet services,'' \url{https://www.ses.com/find-service/commercial-aviation/business-aviation}, last Accessed: 2023.

\bibitem{maritime}
------, ``{SAT-based} maritime internet services,'' \url{https://www.ses.com/find-service/commercial-maritime}, last Accessed: 2023.

\bibitem{fastInternet}
{VIASAT}, ``{SAT-based} fast internet,'' \url{https://www.viasat.com/satellite-internet/}.

\bibitem{disaster}
{SES Luxembourg}, ``{SAT-based} disaster recovery,'' \url{https://www.ses.com/find-service/government/hadr}.

\bibitem{remote}
------, ``{SAT-based} connection of remote communities,'' \url{https://www.ses.com/find-service/telco-mno}.

\bibitem{broadcasting}
{Eutelsat}, ``{SAT-based} dth broadcasting,'' https://www.eutelsat.com/en/satellite-communication-services/broadcasting-solutions.html.

\bibitem{IoT}
------, ``{SAT-based} iot connectivity,'' \url{https://www.eutelsat.com/en/satellite-communications-services.html}.

\bibitem{SWAN}
O.~Michel and E.~Keller, ``Sdn in wide-area networks: A survey,'' in \emph{2017 Fourth International Conference on Software Defined Systems (SDS)}, Valencia, Spain, 2017, pp. 37--42.

\bibitem{electronics13153011}
C.~Fu, B.~Wang, and W.~Wang, ``Software-defined wide area networks (sd-wans): A survey,'' \emph{Electronics}, vol.~13, no.~15, 2024.

\bibitem{al2021migrating}
Y.~Al~Mtawa, A.~Haque, and H.~Lutfiyya, ``Migrating from legacy to software defined networks: A network reliability perspective,'' \emph{IEEE Transactions on Reliability}, vol.~70, no.~4, pp. 1525--1541, 2021.

\bibitem{yungaicela2024misconfiguration}
N.~M. Yungaicela-Naula, V.~Sharma, and S.~Scott-Hayward, ``Misconfiguration in o-ran: Analysis of the impact of ai/ml,'' \emph{Computer Networks}, p. 110455, 2024.

\bibitem{SDVehicles}
D.~F. Blanco, F.~Le~Mouël, T.~Lin, and M.-P. Escudié, ``A comprehensive survey on software as a service (saas) transformation for the automotive systems,'' \emph{IEEE Access}, vol.~11, pp. 73\,688--73\,753, 2023.

\bibitem{Rojas2018}
E.~Rojas, R.~Doriguzzi-Corin, S.~Tamurejo, A.~Beato, A.~Schwabe, K.~Phemius, and C.~Guerrero, ``{Are we ready to drive software-defined networks? A comprehensive survey on management tools and techniques},'' \emph{ACM Computing Surveys}, vol.~51, no.~2, 2018.

\bibitem{NICE}
M.~Canini, D.~Venzano, P.~Pere\v{s}\'{\i}ni, D.~Kosti\'{c}, and J.~Rexford, ``A {NICE} way to test openflow applications,'' in \emph{Proceedings of the 9th USENIX Conference on Networked Systems Design and Implementation}, ser. NSDI'12.\hskip 1em plus 0.5em minus 0.4em\relax San Jose CA, USA: USENIX Association, 2012, p.~10.

\bibitem{TASTE}
D.~Lebrun, S.~Vissicchio, and O.~Bonaventure, ``Towards test-driven software defined networking,'' in \emph{2014 IEEE Network Operations and Management Symposium (NOMS)}, Krakow, Poland, 2014, pp. 1--9.

\bibitem{ATPG}
H.~Zeng, P.~Kazemian, G.~Varghese, and N.~McKeown, ``Automatic test packet generation,'' in \emph{Proceedings of the 8th International Conference on Emerging Networking Experiments and Technologies}, ser. CoNEXT '12.\hskip 1em plus 0.5em minus 0.4em\relax New York, NY, USA: Association for Computing Machinery, 2012, p. 241–252.

\bibitem{BUZZ}
S.~K. Fayaz, T.~Yu, Y.~Tobioka, S.~Chaki, and V.~Sekar, ``Buzz: Testing context-dependent policies in stateful networks,'' in \emph{Proceedings of the 13th Usenix Conference on Networked Systems Design and Implementation}, ser. NSDI'16.\hskip 1em plus 0.5em minus 0.4em\relax Santa Clara CA, USA: USENIX Association, 2016, p. 275–289.

\bibitem{opendaylight}
\BIBentryALTinterwordspacing
{The Linux Foundation}, ``{OpenDaylight a modular open platform for customizing and automating networks of any size and scale.}'' 2023. [Online]. Available: \url{https://www.opendaylight.org/}
\BIBentrySTDinterwordspacing

\bibitem{VERSA}
\BIBentryALTinterwordspacing
V.~Networks, ``Versa secure sd-wan.'' 2023. [Online]. Available: \url{https://versa-networks.com/products/sd-wan/}
\BIBentrySTDinterwordspacing

\bibitem{Lee2020}
S.~Lee, J.~Kim, S.~Woo, C.~Yoon, S.~Scott-Hayward, V.~Yegneswaran, P.~Porras, and S.~Shin, ``A comprehensive security assessment framework for software-defined networks,'' \emph{Computers \& Security}, vol.~91, p. 101720, 2020.

\bibitem{Jero}
S.~Jero, X.~Bu, C.~Nita-Rotaru, H.~Okhravi, R.~Skowyra, and S.~Fahmy, ``Beads: Automated attack discovery in openflow-based sdn systems,'' in \emph{Research in Attacks, Intrusions, and Defenses}, M.~Dacier, M.~Bailey, M.~Polychronakis, and M.~Antonakakis, Eds.\hskip 1em plus 0.5em minus 0.4em\relax Cham: Springer International Publishing, 2017, pp. 311--333.

\bibitem{Bertolino2021}
A.~Bertolino, P.~Braione, G.~{De Angelis}, L.~Gazzola, F.~Kifetew, L.~Mariani, M.~Orr{\`{u}}, M.~Pezz{\`{e}}, R.~Pietrantuono, S.~Russo, and P.~Tonella, ``{A Survey of Field-based Testing Techniques},'' \emph{ACM Computing Surveys}, vol.~54, no.~5, 2021.

\bibitem{NMAP}
G.~F. Lyon, \emph{Nmap Network Scanning: The Official Nmap Project Guide to Network Discovery and Security Scanning}.\hskip 1em plus 0.5em minus 0.4em\relax Sunnyvale, CA, USA: Insecure, 2009, available at http://nmap.org.

\bibitem{IF}
K.~M. Liu Fei~Tony, Ting and Z.~Zhi-Hua, ``Isolation-based anomaly detection,'' \emph{ACM Transactions on Knowledge Discovery from Data (TKDD)}, 2012.

\bibitem{LOF}
M.~M. Breunig, H.-P. Kriegel, R.~T. Ng, and J.~Sander, ``Lof: Identifying density-based local outliers,'' \emph{SIGMOD Rec.}, vol.~29, no.~2, p. 93–104, may 2000.

\bibitem{RamaswamyKNN}
S.~Ramaswamy, R.~Rastogi, and K.~Shim, ``Efficient algorithms for mining outliers from large data sets,'' \emph{SIGMOD Rec.}, vol.~29, no.~2, p. 427–438, may 2000.

\bibitem{King:2014}
R.~S. King, \emph{Cluster Analysis and Data Mining: An Introduction}.\hskip 1em plus 0.5em minus 0.4em\relax USA: Mercury Learning \& Information, 2014.

\bibitem{mcqueen1967smc}
J.~MacQueen, ``Some methods for classification and analysis of multivariate observations,'' in \emph{Proceedings of the 5th Berkeley Symposium on Mathematical Statistics and Probability - Vol. 1}, L.~M. {Le Cam} and J.~Neyman, Eds.\hskip 1em plus 0.5em minus 0.4em\relax University of California Press, Berkeley, CA, USA, 1967, pp. 281--297.

\bibitem{saied2020formal}
W.~Saied and A.~Bouhoula, ``A formal approach for automatic detection and correction of sdn switch misconfigurations,'' in \emph{2020 16th International Conference on Network and Service Management (CNSM)}.\hskip 1em plus 0.5em minus 0.4em\relax Niagara Falls, Canada: IEEE, 2020, pp. 1--5.

\bibitem{saadaoui2019automated}
A.~Sa{\^a}daoui, N.~Ben Youssef Ben~Souayeh, and A.~Bouhoula, ``Automated and optimized formal approach to verify sdn access-control misconfigurations,'' in \emph{Testbeds and Research Infrastructures for the Development of Networks and Communities: 13th EAI International Conference, TridentCom 2018, Shanghai, China, December 1-3, 2018, Proceedings 13}.\hskip 1em plus 0.5em minus 0.4em\relax Springer, 2019, pp. 96--112.

\bibitem{FlowTable_misconfigurations_SDN}
S.~Al-Haj and W.~J. Tolone, ``Flowtable pipeline misconfigurations in software defined networks,'' in \emph{2017 IEEE Conference on Computer Communications Workshops (INFOCOM WKSHPS)}, Atlanta, GA, USA, 2017, pp. 247--252.

\bibitem{pan2022misconfiguration}
H.~Pan, Z.~Li, P.~Zhang, P.~Cui, K.~Salamatian, and G.~Xie, ``Misconfiguration-free compositional sdn for cloud networks,'' \emph{IEEE Transactions on Dependable and Secure Computing}, vol.~20, no.~3, pp. 2484--2499, 2022.

\bibitem{Lebrun2014}
D.~Lebrun, S.~Vissicchio, and O.~Bonaventure, ``Towards test-driven software defined networking,'' in \emph{2014 IEEE Network Operations and Management Symposium (NOMS)}, Krakow, Poland, 2014, pp. 1--9.

\bibitem{manes2019art}
V.~J. Man{\`e}s, H.~Han, C.~Han, S.~K. Cha, M.~Egele, E.~J. Schwartz, and M.~Woo, ``The art, science, and engineering of fuzzing: A survey,'' \emph{IEEE Transactions on Software Engineering}, vol.~47, no.~11, pp. 2312--2331, 2019.

\bibitem{Jero2017}
S.~Jero, X.~Bu, C.~Nita-Rotaru, H.~Okhravi, R.~Skowyra, and S.~Fahmy, ``Beads: Automated attack discovery in openflow-based sdn systems,'' \emph{Lecture Notes in Computer Science (Including Subseries Lecture Notes in Artificial Intelligence and Lecture Notes in Bioinformatics)}, vol. 10453 LNCS, pp. 311--333, 2017.

\bibitem{Fragscapy}
Fragscapy. \url{https://github.com/AMOSSYS/Fragscapy}.

\bibitem{Pham2020}
V.~T. Pham, M.~Bohme, and A.~Roychoudhury, ``Aflnet: A greybox fuzzer for network protocols,'' pp. 460--465, 2020.

\bibitem{Black2021}
C.~Black and S.~Scott-Hayward, \emph{{A Survey on the Verification of Adversarial Data Planes in Software-Defined Networks}}.\hskip 1em plus 0.5em minus 0.4em\relax Association for Computing Machinery, 2021, vol.~1, no.~1.

\bibitem{Scapy}
\BIBentryALTinterwordspacing
Scapy. [Online]. Available: \url{https://scapy.net/}
\BIBentrySTDinterwordspacing

\bibitem{Zeng2012}
H.~Zeng, P.~Kazemian, G.~Varghese, and N.~McKeown, ``Automatic test packet generation,'' in \emph{Proceedings of the 8th International Conference on Emerging Networking Experiments and Technologies}, New York, USA, 2012.

\bibitem{Fayaz2016}
S.~K. Fayaz, T.~Yu, Y.~Tobioka, S.~Chaki, and V.~Sekar, ``Buzz: Testing context-dependent policies in stateful networks,'' in \emph{Proceedings of the 13th USENIX Symposium on Networked Systems Design and Implementation (NSDI'16)}, Santa Clara, CA, 2016, pp. 275--289.

\bibitem{perevsini2015monocle}
P.~Pere{\v{s}}{\'\i}ni, M.~Ku{\'z}niar, and D.~Kosti{\'c}, ``Monocle: Dynamic, fine-grained data plane monitoring,'' in \emph{Proceedings of the 11th ACM Conference on Emerging Networking Experiments and Technologies}, Heidelberg, Germany, 2015, pp. 1--13.

\bibitem{tseng2017sping}
F.-H. Tseng, K.-D. Chang, S.-C. Liao, H.-C. Chao, and V.~C. Leung, ``sping: a user-centred debugging mechanism for software defined networks,'' \emph{IET Networks}, vol.~6, no.~2, pp. 39--46, 2017.

\bibitem{bu2016every}
K.~Bu, X.~Wen, B.~Yang, Y.~Chen, L.~E. Li, and X.~Chen, ``Is every flow on the right track?: Inspect sdn forwarding with rulescope,'' in \emph{IEEE INFOCOM 2016-The 35th Annual IEEE International Conference on Computer Communications}.\hskip 1em plus 0.5em minus 0.4em\relax San Francisco, CA, USA: IEEE, 2016, pp. 1--9.

\bibitem{agarwal2014sdn}
K.~Agarwal, E.~Rozner, C.~Dixon, and J.~Carter, ``Sdn traceroute: Tracing sdn forwarding without changing network behavior,'' in \emph{Proceedings of the third workshop on Hot topics in software defined networking}, Chicago Illinois, USA, 2014, pp. 145--150.

\bibitem{wang2016tool-strace}
Y.~Wang, J.~Bi, and K.~Zhang, ``A tool for tracing network data plane via sdn/openflow,'' \emph{Science China Information Sciences}, vol.~60, no.~2, 2016.

\bibitem{hu2014flowguard}
H.~Hu, W.~Han, G.-J. Ahn, and Z.~Zhao, ``Flowguard: Building robust firewalls for software-defined networks,'' in \emph{Proceedings of the third workshop on Hot topics in software defined networking}, Chicago Illinois, USA, 2014, pp. 97--102.

\bibitem{zeng2014libra}
H.~Zeng, S.~Zhang, F.~Ye, V.~Jeyakumar, M.~Ju, J.~Liu, N.~McKeown, and A.~Vahdat, ``Libra: Divide and conquer to verify forwarding tables in huge networks,'' in \emph{11th USENIX Symposium on Networked Systems Design and Implementation (NSDI 14)}, Seattle, WA, USA, 2014, pp. 87--99.

\bibitem{HSA2012}
P.~Kazemian, G.~Varghese, and N.~McKeown, ``Header space analysis: Static checking for networks,'' in \emph{9th USENIX Symposium on Networked Systems Design and Implementation (NSDI 12)}.\hskip 1em plus 0.5em minus 0.4em\relax San Jose CA, USA: USENIX Association, 2012, pp. 113--126.

\bibitem{fayaz2014testing}
S.~K. Fayaz and V.~Sekar, ``Testing stateful and dynamic data planes with flowtest,'' in \emph{Proceedings of the third workshop on Hot topics in software defined networking}, Chicago Illinois, USA, 2014, pp. 79--84.

\bibitem{Berkhin2006}
P.~Berkhin, \emph{A Survey of Clustering Data Mining Techniques}.\hskip 1em plus 0.5em minus 0.4em\relax Berlin, Heidelberg: Springer Berlin Heidelberg, 2006, pp. 25--71.

\bibitem{TNN:SURVEY:2005}
R.~Xu and D.~Wunsch, ``Survey of clustering algorithms,'' \emph{IEEE Transactions on Neural Networks}, vol.~16, no.~3, pp. 645--678, 2005.

\bibitem{DataClusteringBook:2013}
C.~C. Aggarwal and C.~K. Reddy, \emph{Data Clustering: Algorithms and Applications}, 1st~ed.\hskip 1em plus 0.5em minus 0.4em\relax Chapman \&amp; Hall/CRC, 2013.

\bibitem{Survey:2015}
D.~Xu and Y.~Tian, ``A comprehensive survey of clustering algorithms,'' \emph{Annals of Data Science}, vol.~2, no.~2, pp. 165--193, 2015.

\bibitem{BIRCH}
T.~Zhang, R.~Ramakrishnan, and M.~Livny, ``Birch: An efficient data clustering method for very large databases,'' in \emph{Proceedings of the 1996 ACM SIGMOD International Conference on Management of Data}, ser. SIGMOD '96.\hskip 1em plus 0.5em minus 0.4em\relax New York, NY, USA: Association for Computing Machinery, 1996, p. 103–114.

\bibitem{PAM}
\emph{Partitioning Around Medoids (Program PAM)}.\hskip 1em plus 0.5em minus 0.4em\relax John Wiley \& Sons, Ltd, 1990, ch.~2, pp. 68--125.

\bibitem{DBSCAN}
M.~Ester, H.-P. Kriegel, J.~Sander, and X.~Xu, ``A density-based algorithm for discovering clusters in large spatial databases with noise,'' in \emph{Proceedings of the Second International Conference on Knowledge Discovery and Data Mining}, ser. KDD'96.\hskip 1em plus 0.5em minus 0.4em\relax Portland, Oregon, USA: AAAI Press, 1996, p. 226–231.

\bibitem{OPTICS}
M.~Ankerst, M.~M. Breunig, H.-P. Kriegel, and J.~Sander, ``Optics: Ordering points to identify the clustering structure,'' in \emph{Proceedings of the 1999 ACM SIGMOD International Conference on Management of Data}, ser. SIGMOD '99.\hskip 1em plus 0.5em minus 0.4em\relax New York, NY, USA: Association for Computing Machinery, 1999, p. 49–60.

\bibitem{MeanShift}
D.~Comaniciu and P.~Meer, ``Mean shift: a robust approach toward feature space analysis,'' \emph{IEEE Transactions on Pattern Analysis and Machine Intelligence}, vol.~24, no.~5, pp. 603--619, 2002.

\bibitem{mullner2011modern}
D.~M{\"u}llner, ``Modern hierarchical, agglomerative clustering algorithms,'' \emph{arXiv preprint arXiv:1109.2378}, 2011.

\bibitem{ROUSSEEUW198753}
P.~J. Rousseeuw, ``Silhouettes: A graphical aid to the interpretation and validation of cluster analysis,'' \emph{Journal of Computational and Applied Mathematics}, vol.~20, pp. 53--65, 1987.

\bibitem{survey:anomaly:2009}
V.~Chandola, A.~Banerjee, and V.~Kumar, ``Anomaly detection: A survey,'' \emph{ACM Comput. Surv.}, vol.~41, no.~3, jul 2009.

\bibitem{OutlierSurvey}
A.~Boukerche, L.~Zheng, and O.~Alfandi, ``Outlier detection: Methods, models, and classification,'' \emph{ACM Comput. Surv.}, vol.~53, no.~3, jun 2020.

\bibitem{isolation_forest}
F.~T. Liu, K.~M. Ting, and Z.-H. Zhou, ``Isolation forest,'' in \emph{2008 8th IEEE International Conference on Data Mining}.\hskip 1em plus 0.5em minus 0.4em\relax Pisa, Italy: IEEE, 2008, pp. 413--422.

\bibitem{KnorrKNN}
E.~M. Knorr and R.~T. Ng, ``Algorithms for mining distance-based outliers in large datasets,'' in \emph{Proceedings of the 24rd International Conference on Very Large Data Bases}, ser. VLDB '98.\hskip 1em plus 0.5em minus 0.4em\relax San Francisco, CA, USA: Morgan Kaufmann Publishers Inc., 1998, p. 392–403.

\bibitem{scholkopf2001estimating}
B.~Sch{\"o}lkopf, J.~C. Platt, J.~Shawe-Taylor, A.~J. Smola, and R.~C. Williamson, ``Estimating the support of a high-dimensional distribution,'' \emph{Neural computation}, vol.~13, no.~7, pp. 1443--1471, 2001.

\bibitem{duan2009cluster}
L.~Duan, L.~Xu, Y.~Liu, and J.~Lee, ``Cluster-based outlier detection,'' \emph{Annals of Operations Research}, vol. 168, pp. 151--168, 2009.

\bibitem{hping}
{Salvatore Sanfilippo}, ``{hping},'' \url{https://github.com/antirez/hping}, last Accessed: 2024.

\bibitem{nmapFirewall}
{NMAP team}, ``{NMAP: Bypassing Firewall Rules},'' \url{https://nmap.org/book/firewall-subversion.html}, last Accessed: 2024.

\bibitem{burkov2020machine}
A.~Burkov, \emph{Machine Learning Engineering}.\hskip 1em plus 0.5em minus 0.4em\relax True Positive Incorporated, 2020.

\bibitem{SES}
{SAT}, ``{SAT}, real name hidden for double blind,'' last Accessed: 2023.

\bibitem{Utest}
H.~B. Mann and D.~R. Whitney, ``{On a Test of Whether one of Two Random Variables is Stochastically Larger than the Other},'' \emph{The Annals of Mathematical Statistics}, vol.~18, no.~1, pp. 50 -- 60, 1947.

\bibitem{replicability}
J.~Malik, ``Replication package,'' \url{https://zenodo.org/records/10535145}, last Accessed: 2024.

\bibitem{IPBG}
{Luxembourg National Research Fund}, ``{INSTRUCT - INtegrated Satellite – TeRrestrial Systems for Ubiquitous Beyond 5G CommunicaTions},'' \url{https://instruct-ipbg.uni.lu/}, last Accessed: 2022.

\end{thebibliography}

% \appendix

% \twocolumn[\section{Data used to derive Table~\ref{table:RQ1_versions}}]
% \label{appendix:pvalues}

% \input{tables/RQ1_VDA_Mannwhitney}

%\FIXME{To be added - not done yet to avoid slowing down the compilation of the document}
% \includepdf[pages=-,scale=1,offset=0mm 0]{papers/NGO-ATUA-TOSEM-2021-FINAL.pdf}


\end{document}